\documentclass{article}

\usepackage{blkarray}
\usepackage[applemac]{inputenc}
\usepackage{graphicx}
\usepackage{enumerate}
\usepackage{color}
\usepackage{natbib}
\usepackage{amsfonts}
\usepackage{amsmath}
\usepackage{graphicx}
\usepackage{subfig}
\usepackage{float}
\usepackage[table]{xcolor}
\usepackage{hyperref}
\usepackage{units}
\usepackage{enumitem}

\newtheorem{theorem}{Theorem}
{\end{theorem}\vskip.2cm}
\newtheorem{claim}{Claim}
\newenvironment{cl}{\vskip.2cm\begin{claim}}%
{\end{claim}\vskip.2cm}
\newtheorem{lemma}{Lemma}
{\end{lemma}\vskip.2cm}
\newtheorem{corolla}{Corollary}
{\end{corolla}\vskip.2cm}
\newtheorem{defini}{Definition}
\newenvironment{defi}{\vskip.2cm\begin{defini}}%
{\end{defini}\vskip.2cm}
\newtheorem{proposi}{Proposition}
{\end{proposi}\vskip.1cm}
\newtheorem{prop}{Proposition}
{\end{prop}\vskip.1cm}
\newtheorem{cla}{Claim}
{\end{cla}\vskip.2cm}
\newtheorem{assump}{Assumption}
{\end{assump}\vskip.1cm}
\newtheorem{hypoth}{Assumption}

{\end{hypoth}\vskip.1cm}
\newtheorem{demo}{Proof:}

{\cqfd \end{demo}}
\newtheorem{remark}{Remark}
{\end{remark}\vskip.3cm}

\def\co{{\rm co}\hskip 1pt}

\def\N{{\mathbb{N}}}

\def\proj{{\rm proj}}
\def\proj0{{\rm proj}_{\co M_0}}

\def\R{{\mathbb{R}}}

\makeatletter

\newcommand{\Rmnum}[1]{\expandafter\@slowromancap\romannumeral #1@}
\makeatother

\renewcommand{\thefootnote}{\fnsymbol{footnote}}	

\title{On the Emergence of Scale-free Production Networks}
\author{Stanislao Gualdi\footnote{CentraleSupélec, stanislao.gualdi@gmail.com}
 and Antoine Mandel\footnote{Paris School of Economics, Universit\'{e} Paris I  
Panth\'{e}on-Sorbonne,antoine.mandel@univ-paris1.fr.}\footnote{Gualdi acknowledges the support of  Labex Louis Bachelier (project number ANR 11-LABX-0019). Mandel acknowledges the support of EU FP7 projects SIMPOL and IMPRESSIONS and H2020 project DOLFINS.  We thank Jean-Philippe Bouchaud, Marco Tarzia and Francesco Zamponi for useful discussions as well as the two referees for their insightful remarks that helped us improve the manuscript.}}
\begin{document}

\makeatletter
\renewcommand{\@biblabel}[1]{}
\makeatother
\maketitle

\begin{abstract} \hskip-16pt

We propose a simple dynamical model of the formation of production networks among monopolistically competitive firms. The model subsumes the standard general equilibrium approach à la Arrow-Debreu but displays a wide set of potential dynamic behaviors. It robustly reproduces key stylized facts of firms' demographics. Our main result is that competition between intermediate good producers generically leads to the emergence of scale-free production networks.
\end{abstract}

\renewcommand*{\thefootnote}{\arabic{footnote}}

\section{Introduction}

The scale-free nature of a wide range of socio-economic networks has been 
extensively documented in the recent literature \cite[see 
e.g.][]{barabasi2009,gabaixpower,schweitzer2009}. However, very few contributions have investigated the micro-economic origins of these structures. In this paper, we approach the issue in the context of production networks.
%, whose scale-free nature has received a particular attention in the recent literature \cite[see e.g.][]{Acemoglu,carvalho2014input}. 

We propose a simple model of out-of-equilibrium dynamics that accounts for the endogenous formation of supply relationships.
 The main insight gained from the model is that the emergence of scale invariance is a natural implication of competition  under the two following assumptions: the number of incoming business opportunities for a firm is independent of its size and the rate at which existing consumers may quit grows linearly with the size of the firm. Scale-free structures then balance the speed at which firms grow and shrink. In other words, competition inherently induces the formation of scale-free structures. This results suggests that institutional aspects of economic activity  could  explain empirical findings about the distribution of firms' size \cite[see][]{axtell2001zipf} as well as the emergence of aggregate volatility, which is related to the presence of fat tails in production networks according to \cite{Acemoglu}. 
  
%
%An example of central concern 
%for macro-economics are production networks whose scale invariance has recently 
%been put forward by \cite{Acemoglu,carvalho2014input} as a potentially major driver of 
%macro-economic fluctuations.
%Hence, part of the origins of aggregate volatility have to be searched for in the micro-economic determinants of the formation of production networks. In order to pursue this inquiry, 

  The backbone of our approach is  a model of monopolistic competition on the 
markets for intermediate goods, akin to the one introduced by \cite{ethier1982} 
(on the basis of \cite{dixit1977}) and popularized by the endogenous growth 
literature \cite[see e.g.][]{romer1990}. In this framework, we represent supply 
relationships as the weighted edges of a network and consider 
out-of-equilibrium dynamics in which  (i) demands are made in nominal terms and 
sellers adjust  their prices to balance real supply and nominal 
demand and (ii) firms progressively adjust their production technologies (i.e. the 
network weights) to prevailing market prices.  When the set of relationships is 
fixed (i.e. only the weights of the network can evolve), the identification with 
the underlying  general equilibrium model is perfect in the sense that (i) the 
adjacency matrix of the network is in a one to one correspondence with the 
underlying general-equilibrium economy and (ii) the model  converges to the underlying general 
equilibrium. However, the context of interest for us is the one where the 
technological structure is not fixed a priori and where, the different 
production goods being assumed substitutable, firms can, in the long-run, adjust 
their production technologies/ supply relationships as a function of market prices. Then, we show that the model does not in general 
admit a steady-state, but settles in a dynamic regime where the distribution of firms' size and 
the structure of the production network are scale-free. 

Beyond the emergence of scale-free networks, the model consistently reproduces a rich set of stylized 
facts of firms' demographics: firms' sizes are Zipf distributed, firms' growth rates follow a Laplace distribution, the variance of growth rates decreases with size, and there is a negative correlation between firms' exit rates and age. This set of stylized facts can be reproduced by a relatively parsimonious model (a simple out-of-equilibrium extension of the Arrow-Debreu framework) and for a relatively large range of parameters.

Although there exist micro-foundations for some of these stylized facts in the general equilibrium literature,  they have rarely been jointly analyzed. More importantly, the general equilibrium literature on firms' demographics makes very strong assumptions about the information available to firms, their farsightedness and their rationality. Its results also rely on very specific assumptions about the distribution of exogenous shocks faced by the firms. In this framework, it is rather difficult to understand what are the actual driving forces of firms' demographics. In contrast, we offer a parsimonious model and a simple explanation of the emergence of fat tails as a  consequence of the asymmetric effets of competition on firms of different size.

The remainder of the paper is organized as follows. In section 2, we discuss the relation of our approach to the  literature. In section 3, we give a detailed  description of the model 
and provide a theoretical analysis  of its asymptotic properties. In section 4, we analyze, via numerical simulations, the stability properties of the model and the detailed properties 
of firms' demographics emerging from its dynamics. Section 5 concludes.

\section{Related literature}
 The network perspective on production structures has been introduced in a series of contributions  that investigated the propagation of shocks in economic systems (see in particular \citet{bak,scheinkman1994,weisbuch2007production,BattistonJEDC,Acemoglu}). Our approach builds on this analogy between input-output structures and directed networks, but focuses on the   
 network formation process and the emergence of scale-invariance. In this respect, it is related to the wide literature on the determinants of the distribution of firms' size and more generally  firm demographics.  A first strand of work ranging from the seminal work of \cite{kalecki1945} and \cite{simon1977skew} to more recent contributions such as \cite{bottazzi2006} have approached the determinants of  firms' demographics through ``island-models" in which the growth of each firm is studied in isolation and driven by exogenous shocks. \citep{klepper2006submarkets} provides deep micro-foundations to such approaches by focusing on the development of sub-markets.  It is also worth pointing out that \citep{gabaixzipf} has adopted a similar approach to explain the size distribution of cities. 
 
Firms' demographics have also been investigated within the general equilibrium framework, from a more systemic perspective. The pioneering contributions of  \citet{hopenhayn1992entry} and \cite{ericson1995markov} consider the linkages between firms' production and entry and exit decisions at the industry level. They investigate the optimal organizational response of an industry with respect to productivity shocks  \citep{hopenhayn1992entry} and/or stochastic competition from inside and outside the industry \citep{ericson1995markov}. However, these contributions do not provide a precise characterization of distributional properties. This gap has been filled by a second series of general equilibrium contributions that have put forward  innovation and growth as the key determinants of the inner organization of the industry.  \citet{klette2004innovating} show that the optimal R$\&$D response of firms facing (Poisson distributed) competitive risks and growth opportunities yields Gibrat's law for the growth of firms (i.e. growth rate independent of the size) and a logarithmic size distribution. \citet{luttmer2007selection} considers a model with monopolistically  competitive firms whose productivity grows stochastically over time. He explains the formation of a scale-free distribution of firms' size through the difference between the trends of productivity growth for incumbent and entering firms. 
 \citet{rossi2007establishment} emphasize the role of the accumulation of industry-specific human capital in the emergence of scale-free distribution of firms' size. In particular, they show that scaling is more important in industries with large industry-specific physical capital.   Their results are backed by empirical evidence on the fact that  \emph{``US sectors with larger physical capital shares exhibit significantly more scale dependence in establishment size dynamics and distributions."}
 
Our approach retains the dynamic perspective of the ``island-based" literature and the systemic perspective of the general-equilibrium literature. It is closer to the strand of literature à la \cite{hopenhayn1992entry} because it focuses on competition \emph{per se} and does not require the introduction of aggregate growth to explain a wide range of stylized facts. The empirical results we obtain are nevertheless very similar. In particular, our emphasis on the role of competition is perfectly consistent with the observation made by  \citet{rossi2007establishment} that  scale dependence increases with fixed costs. 

 The innovation our model brings to the literature is its relative parsimony and the reduced role of stochastic shocks in the emergence of scale invariance. Yet, our most important contribution is to provide a model of formation of production networks whereas in the existing literature the extent of direct interactions between firms is fairly limited (general equilibrium linkages generally occur only through final demand).  In this respect, the  paper contributes to the literature on the formation of 
socio-economic networks \cite[see e.g.][]{jackson2008} by providing 
micro-foundations for the emergence of scale-free networks which have been 
largely lacking in this literature, except for notably 
\cite{jackson2007meeting}. The recent contribution of  \cite{carvalho2014input} that applies the model of Jackson and Rogers to the formation of production networks is hence very closely related to the present contribution. However, our approach also offers a bridge to the general equilibrium literature. 

\section{The Model} \label{sec-model}

\subsection{The general equilibrium framework}

We consider an economy consisting of a finite set of monopolistically 
competitive firms producing differentiated goods and of a representative household. We denote  the 
set of firms by $M=\{1,\cdots,m\}$, the representative household by the index $0$ and the set of agents  by $N=\{0,\cdots,m\}$. 

The representative household supplies a constant quantity of labor (normalized 
to $1$) and has preferences represented  by a Cobb-Douglas utility function of 
the form $u(x_1,\cdots,x_m)=\prod_{i=1}^m x_i^{\alpha_{0,i}}.$ He shall hence spend
his income on each good $i \in M$ proportionally to $\alpha_{0,i}$ (we assume 
that for all $i \in M, \alpha_{0,i}>0,$ so that the household consumes a 
positive quantity of each and every good).

With respect to firms, our central concern is the endogenous formation of supply relationships. To assume away any exogenous determinism in this respect, we place ourselves in a setting where there is no a priori distinction between potential intermediate goods. 
That is, as in standard models of monopolistic competition on the intermediate goods 
markets \cite[see][]{ethier1982,romer1990}, we consider that each good can be used interchangeably in the production process. More precisely, we consider that the production possibilities of firm $i$ are given by a C.E.S production function of the form:
% *** (todo mention the possibility of weights) ***

 \begin{equation} 
 g_i(x_0,(x_{i_j})_{j=1,\cdots,n_i})= x_0^{\alpha} (\sum_{j=1}^{n_i} x_{i_j}^{\theta})^{ 
\nicefrac{(1-\alpha)}{\theta}}  
\label{prodGene} 
\end{equation}
where $\alpha \in (0,1)$ is the (nominal) share of labor in the input mix, $\nicefrac{1}{(1-\theta)}$ is the elasticity of substitution, $x_0 \in \R_+$ is the quantity of labor used, $n_i \in \N$ the number of intermediate goods combined and $x_{i_j}$ the quantity of the $j$th input used by firm $i$  in its production process\footnote{Note that $n_i$ refers here to the number of goods combined not to any indexing of the goods. Equation \ref{prodSpec} provides a more precise characterization.}.  

\begin{remark}\label{ces} We assume throughout the paper that  $\theta \in [0,1],$  that is, inputs are gross substitute. This implies in particular that productivity grows with the number of  inputs/suppliers combined. This feature is also at the core of the infra-marginal approach to economic growth \cite[see][]{yang1991} and of Adam Smith's original description of the effects of the division of labor.  Also note that $\theta=0$ corresponds to the case of a Cobb-Douglas production function and $\theta=1$ to  a linear production function, i.e. perfectly substitutable inputs.\footnote{
As a matter of fact, we have also conducted numerical experiments for $\theta \in (-\infty,0).$ The inner workings of the model are similar to the one presented below but for the fact that low out-degree firms then are the most productive ones. Most of the results are symmetric except those that rely on the fact that there is no absolute bound on the productivity of firms  when inputs are substitutable (i.e. each firm can potentially face a more competitive peer). Indeed, when input are gross complements, firms with a single supplier have maximal productivity. 
} 

\end{remark}

As such, this model is incomplete. In order to fully determine the micro-economic choices of the agents in terms of production or consumption, we have to introduce additional assumptions on the structure of interactions, i.e. we have to specify the production network. This network can be characterized by an adjacency matrix, $A=(a_{i,j})_{i,j \in M},$ such that $a_{i,j}=1$ if $j$ is a supplier of $i$ and $a_{i,j}=0$ otherwise. Letting $S_i(A):=\{ j \in M \mid a_{i,j}=1\}$ denote the set of suppliers of firm $i,$ the production function of firm $i$ is then further specialized into:  

\begin{equation} 
f_i(x_0,(x_j)_{j \in S_i(A)})= x_0^{\alpha} (\sum_{j \in S_i(A)} x_j^{\theta})^{ 
\nicefrac{(1-\alpha)}{\theta}} 
\label{prodSpec} 
\end{equation}
Given such a production network $A,$  one can  close the model ``à la Arrow-Debreu" by associating to the production network $A,$  an economy $\mathcal{E}(A)$ and a notion of general equilibrium. The  economy $\mathcal{E}(A)$ is defined as follows.

\begin{defi}
The general equilibrium economy $\mathcal{E}(A)$ is defined by:
\begin{itemize}
\item A representative household supplying one unit of labor and having Cobb-Douglas preferences $u(x_1,\cdots,x_m)=\prod_{i=1}^m x_i^{\alpha_{0,i}}.$
\item A set of firms $M$ with production functions of the form
\begin{equation} 
 f_i(x_0,(x_j)_{j=1,\cdots,n_i})= x_0^{\alpha} (\sum_{j=1}^{n_i} x_j^{\theta})^{ 
\nicefrac{(1-\alpha)}{\theta}}  
\end{equation}
\item A production network  $A$ consistent with equation  $(\ref{prodGene})$ in the sense that for all $i \in M,$ $\sum_{j=1}^m a_{i,j}=n_i.$
\end{itemize}
\end{defi}
As we shall consider in the following that firms are monopolistic competitors, the notion of general equilibrium we introduce entails monopolistic mark-ups. Namely:
\begin{defi} A general equilibrium with monopolistic mark-up $\lambda$ of the economy $\mathcal{E}(A)$ is a 
collection of prices $(p^*_0,\cdots,p^*_n) \in \R^M_+$, production levels 
$(q^*_0,\cdots,q^*_n) \in \R^M_+$ and commodity flows $(x^*_{i,j})_{i,j=0\cdots 
n} \in \R^{M\times M}_+$ such that:
\begin{enumerate}
\item Markets clear. That is for all $i \in M,$ one has  $$q^*_i = \sum_{j=1}^M 
x^*_{i,j}.$$
\item The representative consumer maximizes his utility. That is $(q^*_0,(x^*_{0,j})_{j=1,\cdots,n})$
 is a solution to 
$$\left\{\begin{array}{c} \text{max } u_i((x_{0,j})_{j=1,\cdots,n}) \\  \\
\text{s.t }  \sum_{j=1}^n 
p^*_j x^*_{0,j} \leq 1\end{array}\right.$$
(with the price of labor normalized to $1$)
\item Production costs are minimized. That is for all $i \in M,$ 
$(x^*_{i,j})_{j=0\cdots n}$ is the solution to 
$$\left\{\begin{array}{cc} \text{min} & \sum_{j \in S_i(A)} p^*_j x_j  \\ 
\text{s.t} & f_i(x_j)\geq q^*_i \end{array}\right.$$
\item Prices are set as a mark-up over production costs at rate 
$\dfrac{\lambda}{1-\lambda}.$ That is one has for all $i \in N:$
$$p^*_i  =(1+ \dfrac{\lambda}{1-\lambda})\dfrac{\sum_{j \in S_i(A)} p^*_j 
x^*_{i,j}}{q^*_i}$$
\end{enumerate}
\label{monequi}
\end{defi}

\begin{remark} \label{remequi}
   In a setting with constant returns to scale, profits are zero at a competitive equilibrium. Hence, for $\lambda=0,$ general equilibrium with monopolistic mark-ups coincide with competitive equilibria where firms maximize profits, that is, where conditions $3.$ and $4.$ are replaced by: 
\begin{enumerate}
\item[3'.] For all $i \in M,$ 
$(q^*_i,(x^*_{i,j})_{j \in S_i(A)})$ is a solution to 
$$\left\{\begin{array}{c} \text{max }  p^*_i q_i- \sum_{j \in S_i(A)} p^*_j x_{i,j}  \\  \\
\text{s.t }  f_i((x_{i,j})_{j \in S_i(A)})\geq q_i \end{array}\right.$$
\end{enumerate}

\end{remark}

\subsection{Dynamics}

At a general equilibrium, supply relationships are considered  fixed a priori. The central concern of this paper is the evolution of these relationships, i.e. the formation of production networks. The existing literature is relatively silent on the topic: in the endogenous growth literature à la \citet{romer1990} and in the general equilibrium approach to firm dynamics à la \citep{luttmer2007selection} or \citep{rossi2007establishment}, the production structure is evolving but there are no direct interactions between firms: all the linkages go through final demand.  Our approach will be akin to the one of the out-of-equilibrium literature \cite[see e.g.][]{arrow1959stability,fisher1989disequilibrium,arthur2006out} but beyond the adjustment of prices and quantities, we will also investigate the evolution of the production structure.

 We assume time is discrete and indexed by $t \in \N.$ Every period, the state of agent $i$ is determined by the wealth it holds $w_i^t \in \R_+,$ the stock of output it has produced $q_i^t \in \R_+,$ the price it sets for its output $p_i^t \in \R_+$ and the input shares $(\alpha_i^t)_{i \in N} \in \R_+^N$ it chooses.  Moreover, at random times, each agent is given the opportunity to adapt its supply relationships to the prevailing market conditions. This determines the evolution of the production network over time, whose adjacency matrix in period $t$ is denoted by  $A^t$ . 

\subsubsection{Out-of-equilibirum dynamics}  \label{outdyn}

More precisely, during each period $t \in \N,$ the following sequence of events takes place:

\begin{enumerate}
\item Each agent $i \in N$ receives the nominal demand $\sum_{j \in N} \alpha^t_{i,j} w^t_j.$ 
\item Agents adjust their prices frictionally towards their market-clearing values according to: 
\begin{equation}p_i^t=\tau_p \overline{p}_i^t+ 
(1-\tau_p) p^{t-1}_i \label{pricefric}\end{equation}
where $\tau_p \in[0,1]$ is a parameter measuring the speed of price adjustment and, given the nominal demand  $\sum_{j \in N} \alpha^t_{i,j} 
w^t_j$ and the product stock $q_i^t,$ $\overline{p}_i^t$ is the market clearing price for firm $i,$ that is: \begin{equation}\overline{p}_i^t=\dfrac{\sum_{j \in 
N} \alpha^t_{i,j} w^t_j}{q_i^t}. \label{price}\end{equation} 

\item  Whenever $\tau_p<1,$ markets do not clear (except if the system is at a stationary equilibrium). In case of excess demand, we assume that clients are rationed proportionally to their demand. In case of excess supply, we assume that the amount 
 $\overline{q}^t_i:=\nicefrac{\sum_{j \in N} \alpha^t_{i,j} 
w^t_j}{p_i^t}$ is actually sold and that the rest of the output is stored as inventory\footnote{The household does not carry an inventory. Equation \ref{prod} is  modified accordingly in this case.}. Together with production occurring on the basis of purchased inputs, this yields the following evolution of the product stock:

\begin{equation}q^{t+1}_i=q^t_i-\overline{q}^t_i + 
f_i(\dfrac{\alpha^t_{0,i}w^t_{i}}{p_0^t},(\dfrac{\alpha^t_{j,i}w^t_{i}}{p_j^t})_{j 
\in S_i(A^t)}) \label{prod}\end{equation}

Note that in the case where  $\tau_p=1,$ markets always clear (one  has  $\overline{q}^t_i = q^t_i$) and equation  $(\ref{prod})$ reduces to 
\begin{equation}q^{t+1}_i= 
f_i(\dfrac{\alpha^t_{0,i}w^t_{i}}{p_0^t},(\dfrac{\alpha^t_{j,i}w^t_{i}}{p_j^t})_{j 
\in S_i(A^t)}) \label{prodbis}\end{equation}

\item As for the  evolution of agents' wealth, it is determined on the one hand 
by their purchases of inputs and  their sales of output. On the other hand, we 
assume that the firm  sets its expenses for next period at $(1-\lambda)$ times its current 
revenues and distributes the rest as dividends to the representative household. One therefore has: 
\begin{equation}\forall i \in M,\ w^{t+1}_i=(1-\lambda)\overline{q}^t_i 
p_i^t\label{evfirm}\end{equation}
\begin{equation}w^{t+1}_0= {q}_0^t p_0^t +\lambda \sum_{i \in M}\overline{q}^t_i 
p_i^t\label{evhous} \end{equation}

Note that equation $(\ref{evfirm})$ can be interpreted as assuming that firms have myopic expectations about their nominal demand (i.e. they assume they will face the same nominal demand next period) and target a fixed profit/dividend share $\lambda \in (0,1).$ 

\item As for the evolution of input shares, agents adjust frictionally their input combinations towards the cost-minimizing value according to: 
\begin{equation} \alpha_i^{t+1}=\tau_{w} \overline{\alpha}_i^{t} +(1-\tau_{w}) 
\alpha_i^t \label{optweightfric}\end{equation}
where $\tau_{w} \in[0,1]$ measures the speed of technological adjustment and $ \overline{\alpha}_i^{t} \in \R^M$ denotes the optimal input weights for firm $i$  given prevailing prices. Those weights are defined as the solution to the following optimization problem: 

\begin{equation}\left\{\begin{array}{cc} \text{max} & 
f_i(\dfrac{\alpha_0,i}{p_0^t},(\dfrac{\alpha_j,i}{p_j^t})_{j \in S_i(A^t)})  \\ 
\text{s.t.} & \sum_{j \in S_i(A^t)} \alpha_{j,i}=1\end{array}\right. 
\label{optweight}\end{equation}
\end{enumerate}

\subsubsection{Evolution of the production network} \label{netdyn}

This first sequence of operations defines out-of-equilibrium dynamics in the general equilibrium economy $\mathcal{E}(A),$ for a given production network $A$. They are akin to the dynamics considered in the literature on the stability of general equilibrium in the Arrow-Debreu framework  \cite[e.g.][]{arrow1959stability,fisher1989disequilibrium,arthur2006out}. As we let the production network evolve, we leave the Arrow-Debreu framework {\it sensu stricto}, and enter the realm of industrial dynamics. In this extended setting, our central concern is to determine if the emergence of empirical regularities, such as scale-free networks, can be explained by the micro-economic decisions of firms. 

For sake of parsimony, and to disentangle the effects of competition from those of technological change, we place ourselves in a setting where there is no aggregate growth in productivity nor output (in the long-term). Therefore, we consider that there can not be entry above the maximal number of firms $m$ and that  the total number of suppliers of each firm  is fixed so that the productivity of a firm cannot grow endogenously (following Remark \ref{ces}, productivity is related to the number of suppliers). The sole driver of the evolution of the network is (monopolistic) competition that we represent by assuming firms can switch suppliers one for one. In other words, we consider that the out-degree distribution of the production network is fixed and focus on the evolution of the in-degree distribution.
 
The detailed evolution of the network is determined as follows. At the end of every period,
each firm independently receives the opportunity to change one of its suppliers with probability 
$\rho_{chg} \in [0,1].$ If this opportunity materializes for firm $i$ in 
period $t,$ it selects randomly one of its suppliers 
$\overline{j}_{i}$ and another random firm $j$ among those to  which it is not 
already connected. It then shifts its connection from firm $\overline{j}_{i}$  
to firm $j$ if and only if the price of $j$ is less than the one of 
$\overline{j}_{i}.$ In other words, the adjacency matrix $A^t$ evolves according 
to:
\begin{equation} \begin{array}{c}  a_{i,\overline{j}_i}^{t+1}= 
\left\{\begin{array}{cc} 1 &  \text{if $p^t_{\overline{j}_i} \leq p^t_{j}$  }  
\\0 & \text{otherwise} \end{array}\right. \\  \\ \ a_{i,j}^{t+1}= 
1-a_{i,\overline{j}_i}^{t+1}  \end{array} \label{netchange}\end{equation}
The actual weight of the new connection is then determined according to an average
over other suppliers' weights.

Finally, the possibility for a firm to lose connections implies that it can eventually be 
driven out of the market. Indeed, we consider that a firm that has lost all its 
connections towards other firms exits the market. To sustain competition in the 
economy, we assume that those exits are compensated by entries of new firms 
according to the following process. Every period, each (potential) firm that is 
out of the market independently enters with probability $p_{new}.$ When 
entering, the firm is endowed with the following characteristics:
\begin{itemize}

 \item The number of suppliers is drawn from a binomial distribution $B(p,n)$. 
The success probability
 $p$ is adjusted in order to preserve the mean degree in the network in the 
 long-run.\footnote{In order to preserve the total number of links in the network we compute the average out-degree of inactive firms $\bar{k}_I$ and
 set the success probability in the binomial to $p=\frac{\bar{k}_I}{n}(1+10\frac{L_0-L_t}{L_t})$ where $L_t$ is the total number of links at time $t$.
 As long as the total number of links does not steadily grow or decline, our  results are independent of this particular implementation.}
 \item The price is initially set equal to the average price in the economy. 
 \item Each firm in the economy rewires to the newly created firm independently 
with probability $\bar{k}/m,$ where $\bar{k}$ is the average number of
 clients at time $0.$
 \item The wealth of the firm is set equal 
to the average wealth of other firms and its initial product stock is empty.\footnote{To ensure conservation of money in the long term, the initial wealth of the firm is in practice considered as a loan that the firm has to reimburse before it can pay any dividend.}  
\end{itemize}

\subsection{Asymptotic analysis}

In section \ref{simu}, we shall explore in depth the dynamic behavior of the model via simulations. Yet, we can characterize key mechanisms at play in the model through an analytical characterization of its asymptotic properties. 

\subsubsection{Asymptotics for a fixed network}

Let us first characterize the steady-state of the out-of-equilibrium dynamics defined in subsection $\ref{outdyn}$ when the production network $A$ is fixed. At a steady state, equation $\ref{prod}$ implies market-clearing and equation $\ref{optweightfric}$ implies the minimization of costs. Therefore one clearly has:
\begin{prop} Let us consider the production network $A$ is fixed and out-of-equilibrium dynamics in the economy $\mathcal{E}(A)$ are defined as in subsection \ref{outdyn}. Then, the steady states of the dynamics are general equilibria of the economy $\mathcal{E}(A)$ with monopolistic mark-up $\lambda$. Moreover, if $\lambda=0,$ these also coincide with competitive equilibria\footnote{See Remark $\ref{remequi}$}.
\label{propfix}\end{prop}

These results underline the fact that our model formally subsumes the standard general equilibrium framework when the network is considered as fixed. Note however that, as illustrated in the next section, the dynamics of the model do not necessarily yield convergence to the steady state/ general equilibrium and that a rich taxonomy of asymptotic behavior can be observed at limit cycles. 
 
\begin{remark} \label{redux}
A direct corollary of this result is that if we let the network structure evolve according to  $\ref{netdyn},$ the model will have the same steady-states as the simplified equilibrium model in which every period:
\begin{itemize} 
\item An equilibrium with monopolistic mark-up $\lambda$ is computed given the structure of the production network.
\item The production network evolves as described in subsection $\ref{netdyn}.$ 
\end{itemize}

\noindent Note, however, that the inclusion of out-of-equilibrium dynamics is crucial for the stability analysis and the detailed investigation of firms' demographics performed below. 
\end{remark}
 
 \subsubsection{Asymptotics with an evolving network}
 
 We now turn to the investigation of asymptotic properties of the model when the production structure evolves  according to the dynamics defined in subsection $\ref{netdyn}.$ 
In this extended setting, our central concern is the characterization of the asymptotic properties of the production network. In particular we are interested
 in the potential emergence of fat tails in the degree distribution, whose presence  has been put forward as a salient feature of socio-economic networks \cite[see e.g.][]{barabasi1999} and, more recently, as a key feature in the build-up of macro-economic volatility \cite[see][]{Acemoglu}.  
 
In our setting, the evolution of the production network is driven by competition between firms. In turn, the primitive determinant of a firm's competitiveness is its production technology, 
which is completely characterized by the number of suppliers/inputs combined (see equation $\ref{prodGene}$). 
Therefore, the number of suppliers of a firm $i,$ $n_i,$ should be the main determinant of its position in the network. 
 In this respect, let us recall that the number of suppliers of a firm is fixed throughout: a firm can change the identity of its suppliers, not their number. 
 In other words, the out-degrees in the network are fixed and we are investigating the evolution of the in-degrees. 

 \subsubsection{Steady-states with an evolving network}

We shall first characterize the structure of the network at a steady-state of the dynamics (given in subsections  $\ref{outdyn}$ and  $\ref{netdyn}$), although it will turn out these steady states are generally unstable.  

The structure of a steady-state of the network is constrained by two conditions. First, links shall be prioritarily directed towards the most competitive firms. Second, the competitiveness of firms increases with the number of suppliers (see remark \ref{ces}). Hence, at a steady state, one shall observe a hierarchical structure for which incoming links are directed first and foremost towards more competitive (high out-degree) firms. Namely, one has:
\begin{prop} If $\overline{A}=(\overline{a}_{i,j})_{i,j \in M}$ is a steady-state production network, then one has for all $i,j \in M:$
\begin{equation} n_j<n_i \Rightarrow [\forall h \in M  \ \overline{a}_{h,j}=1 \Rightarrow \overline{a}_{h,i}=1]\end{equation}
\label{propns}\end{prop}
\begin{demo} See Appendix $\ref{proofns}.$
\end{demo}

Proposition $\ref{propns}$ implies that a firm can be linked to a firm with $k$ suppliers only if it is already linked with all the firms that have more than $k$ suppliers (i.e. the more competitive firms). 
This means that steady state networks shall have a very specific structure: firms are grouped in clusters according to their out-degree (number of suppliers) and there are links from a firm towards a cluster only if it is already linked to all
firms in higher out-degree clusters (see Appendix $\ref{stnet}$ for a detailed description). 

\begin{remark}
This structure is reminiscent of nested-split graphs  \cite[see][]{mahadev,konig2012, konig2014}. The only difference being the fact that in our setting a cluster is not necessarily linked to all clusters with higher out-degrees than its own.
Indeed, clusters are associated to out-degrees and, by definition, the lower the out-degree of a cluster the fewer connections it has.
\label{remark3}
\end{remark}

Another consequence of Proposition $\ref{propns}$ is that at a steady-state only the firms with the maximal number of suppliers actually have 
consumers. In particular, for each firm to actually have at least one consumer at the steady-state, there must exist firms with a very large number of suppliers for the whole set of potential suppliers to be spanned.  
 If this condition is not met, the least productive firms have no consumers and therefore, as described in subsection $\ref{netdyn},$ they must exit the market. However, exit contradicts the very fact that the system is at a steady-state. Hence, unless there are firms with very large number of suppliers, the system does not have a steady-state.  
 
As a matter of fact, convergence to a steady-state is not observed in simulations (see below) unless the system is initialized in a very peculiar state (e.g. by letting all the firms exactly have the same number of suppliers).  On the contrary, we generically observe sustained entry and exit, growth and decline of firms.  It therefore seems that steady states are of little relevance to characterize the asymptotic behavior of the model. 
 
 \subsubsection{Stable degree distributions}
 
Meso-level distributional characteristics can be asymptotically stable even in absence of convergence at the micro-level \cite[see][for an extended discussion of the issue]{aoki2011}. In our setting, the (in-)degree distribution of the production network can be stable without the network itself being at a microscopic steady state. For example, the degree distribution stays unchanged if a firm of degree $k$ loses a link to a firm of degree $(k-1):$ the firm formerly of degree $k$ becomes of degree $(k-1)$ while the firm formerly of degree $k-1$ becomes of degree $k.$   

In the following, we provide a characterization of the stable degree distribution in a stylized version of the model, which highlights how competition shapes the structure of the production network. In a nutshell, there are two effects at play. On the one hand the number of incoming business opportunities for a firm is independent of its size. On the other hand, the larger a firm is, the more it is exposed to competition (i.e. the more likely it is that one of its customer considers switching). Then, at a statistical equilibrium, the degree distribution must be scale-free in order to balance the flow of incoming and outgoing links.  
 
 In order to clarify this argument, let us consider a simplified version of the model in which the rewiring probability $\rho_{chg}$ is chosen small enough so that there is a single link swap per period\footnote{This is of course an approximation but if $\rho_{chg}$ is small enough, the probability that there is two link swaps in a given period is negligible with respect to the probability that there is only one.}. Then each firm can win or lose at most one link per period  and accordingly each degree class can win or lose at most one firm per period. Therefore a degree distribution $(P_k)_{k \in \N}$ is stationary if  the probabilities for a firm of degree $k$ to lose and to gain a link compensate each other. 

In order to approximate the probability $\rho_k$ for a firm of degree $k$ to gain a link, let us first remark that the number of incoming business opportunities for a firm $j$ is independent of its size (because the switching firm chooses its tentative new supplier uniformly at random). Hence the probability for a firm to have the opportunity to catch an incoming link is $\nicefrac{1}{m}.$ According to Equation $\ref{netchange},$ the link will actually be caught if the incumbent supplier is more expensive than firm $j.$ Now, if one assumes that the competitiveness of the incumbent supplier is independent of the identity of the switching firm, one can consider that the switching link is drawn uniformly at random among links. Then, the probability for firm $j$ to actually catch the link, given it has been drawn as a challenger, is given by the share  $\psi_j  \in [0,1]$ of existing links towards more expensive firms. Summing-up, one can write the probability for firms of degree $k$ to gain a link as:

\begin{equation} \rho_k= \sum_{\{j\mid d_j=k \}}\dfrac{1}{m}\psi_j \label{totgain0} \end{equation}
where $d_j$ denotes the degree of firm $j.$ 
After introducing $n_k$  the number of firms of degree $k,$ the equation can be rewritten as follows: 
\begin{equation} \rho_k= \dfrac{n_k}{m}\sum_{\{j\mid d_j=k \}}\dfrac{\psi_j}{n_k} =P_k \psi_k\label{totgain} \end{equation}
where $ \psi_k$ is the average share of links towards more expensive firms for firms of degree $k.$

In order to approximate the probability $\mu_k$  for a firm of degree $k$ to lose a link, let us remark that  the larger a firm is (i.e. the higher its in-degree), the more it is exposed to competition. More precisely, the probability that a firm's consumer considers switching suppliers is proportional to the number of consumers that the firm has (i.e. to its in-degree). Hence the probability for a firm $j$ of degree $d_j$ to be at risk of losing a consumer is $\nicefrac{d_j}{m \tilde{d}}$ where $\tilde{d}$ is the  mean degree in the network (and hence $m\tilde{d}$ is the total number of links in the network). Now, firm $j$ will actually lose the link if the challenger firm, which is selected uniformly at random, is more competitive. Hence the probability for firm $j$  to actually lose a link, given it has been drawn as an incumbent, is given by the share  $\phi_j  \in [0,1]$ of firms that are cheaper than $j.$ Summing-up, one can write the probability for firms of degree $k$ to lose a link as: 

\begin{equation} \mu_k= \sum_{\{j\mid d_j=k \}}\dfrac{d_{j}}{m\tilde{d}}\phi_j= \sum_{\{j\mid d_j=k \}}\dfrac{k}{m\tilde{d}}\phi_j \label{totloss0} \end{equation}
Or equivalently
\begin{equation} \mu_k=\dfrac{1}{\tilde{d}}k \dfrac{n_k}{m}\sum_{\{j\mid d_j=k \}} \dfrac{\phi_j}{n_k}=\dfrac{1}{\tilde{d}}k P_k \phi_k \label{totloss} \end{equation}
where $\phi_k$ is the average position in the price ordering of firms of degree $k.$

Contrasting equations  (\ref{totgain}) and (\ref{totloss}), one remarks that the competitive pressure faced by a firm increases linearly with its degree, while the number of incoming business opportunities  are independent of the current size/degree. Namely, the ratio between the flows of incoming and outgoing links is given by:
   \begin{equation}\dfrac{\mu_k}{\rho_{k}}=\dfrac{k}{\tilde{d}} \dfrac{\phi_k }{\psi_{k}} \label{necme},\end{equation} 

As emphasized above, a sufficient condition for a  distribution to be stationary is then to balance the flow of incoming and outgoing links. Namely if $(P_k)_{k \in \N}$ is a stationary distribution, $F_k$ its cumulative distribution and $\rho_k$ and $\mu_k$ the associated transition rates, one shall have $\rho_k \simeq \mu_k.$ Now, at a stationary distribution,  
the competitiveness of firms shall increase with the degree and hence firms shall only lose incoming links towards higher degree firms. In other words, the share of firms that are more expensive than firms of degree $k,$ $\phi_k,$ is the tail of the degree distribution, $1-F_k.$  Hence,  one has: 
\begin{equation} \mu_k \simeq \dfrac{1}{\tilde{d}}k P_k (1-F_k).\end{equation}
and therefore
\begin{equation} P_k \psi_k \simeq \dfrac{1}{\tilde{d}}k P_k (1-F_k).\end{equation}
 As moreover $\psi_k,$ being a probability, satisfies $\lim_{k \rightarrow + \infty} \psi_k=1,$  it must be that:
\begin{equation}  1-\overline{F}_k \sim_{k \rightarrow + \infty}   \dfrac{\tilde{d}}{k}. \end{equation} 

 That is the degree distribution of the production network asymptotically follows Zipf's law. Formally, one has: 
 
\begin{prop}  Assume $\rho_k$ and $\mu_k$ are  transition rates such that Equation (\ref{necme}) holds and let $(\overline{P}_k)_{k \in \N}$ be an associated stationary distribution. Then, one has:

\begin{equation}   \overline{F}_k:=\sum_{\ell=1}^k \overline{P}_{\ell} \sim_{+ \infty}  1-\dfrac{\tilde{d}}{k} \end{equation}
\label{propme}
\end{prop}

\noindent Several remarks are in order about Proposition  $\ref{propme}.$
\begin{enumerate}
\item  The proposition focuses on the formation of the tail of the degree distribution. The asymptotic nature of the result implies that 
one abstracts away from the finite nature of the numerical model and considers a limit model with a countably infinite number of firms. Note in this respect that condition $(\ref{necme})$ is stated independently of the number of firms. 
\item We obtain asymptotic convergence to a scale-free distribution with logarithmically diverging mean while the mean is, by construction, finite in the original model. This discrepancy might be explained by the fact that 
our analytical approximation overestimates the probability of receiving an incoming link as it does not account for the entry and exit of firms nor for the fact that firms must choose distinct suppliers. Nevertheless, the analytical result is  clearly in line with the results of numerical experiments (see below), but for a small upward bias in the exponent of the distribution.
  \item The existence and the characterization of the stationary degree distribution are independent of the elasticity of substitution and of most of the other parameters of the model.
\item A complete characterization of the class of distribution obeying a master equation with transition rates given by equations $(\ref{totgain})$ and  $(\ref{totloss})$ is beyond the scope of this paper but  preliminary results suggest that
they might yield a rich set of scale-free structures.

\end{enumerate}

\noindent Proposition $\ref{propme}$ highlights the fact that competition shapes the structure of the production network. The structure must be consistent with the speeds at which  firms grow (at the expense of less competitive peers) and shrink (in favor of more competitive peers). Fat-tails in the degree distribution hence emerge because of two basic facts about the ``economy" of suppliers' switches.  On the one hand,  the number of incoming business opportunities for a firm is independent of its size, i.e. firms gain link at a constant rate. On the other hand, the larger a firm is (i.e. the higher its in-degree), the more it is exposed to competition. Indeed, the rate at which existing consumers may quit grows linearly with the size of the firm, i.e. firms lose links proportionally to their degree. Then, at a statistical equilibrium, the degree/size distribution must be scale-free in order to balance the flow of incoming and outgoing links.  

 The  process can be seen as a from of inverted preferential attachment  \cite[see][]{barabasi1999} where asymptotically large firms lose connections proportionally to their degree (whereas in the standard preferential attachment model finite-size nodes 
gain new links proportionally to their degree).
\section{Numerical Analysis}
\label{simu}

Numerical simulations allow us to investigate the dynamic stability of the model and to compare its distributional properties with empirical data on firm demographics. We report below the result of an extensive series of simulations for a model with $M=10000$ firms (and $M=2000$ firms in the simplest case with $\rho_{chg}=0$) and representative samples of parameters.  

\subsection{Stability}

\subsubsection{General Equilibrium}\label{gestab}

Proceeding stepwise as in the preceding section, we first consider the case where the adjacency structure is fixed and therefore focus on the stability of a general equilibrium with monopolistic  mark-up $\lambda$  for the out-of-equilibrium dynamics introduced in subsection $\ref{outdyn}.$ This problem is very similar to that of the stability of general equilibrium, which has been extensively studied in the literature \cite[see][for a recent survey]{fisher2011stability}. However, in our framework, behavior is much more local/decentralized and boundedly rational than generally is in the general equilibrium literature (firms have myopic expectations about their demand and update their prices in an adaptive manner).  Therefore, it is extremely difficult to apply the standard apparatus of differential calculus to investigate stability issues. However, simulations provide very clearcut results. 

\begin{figure}[h]
\begin{center}
\includegraphics[width=0.5\linewidth]{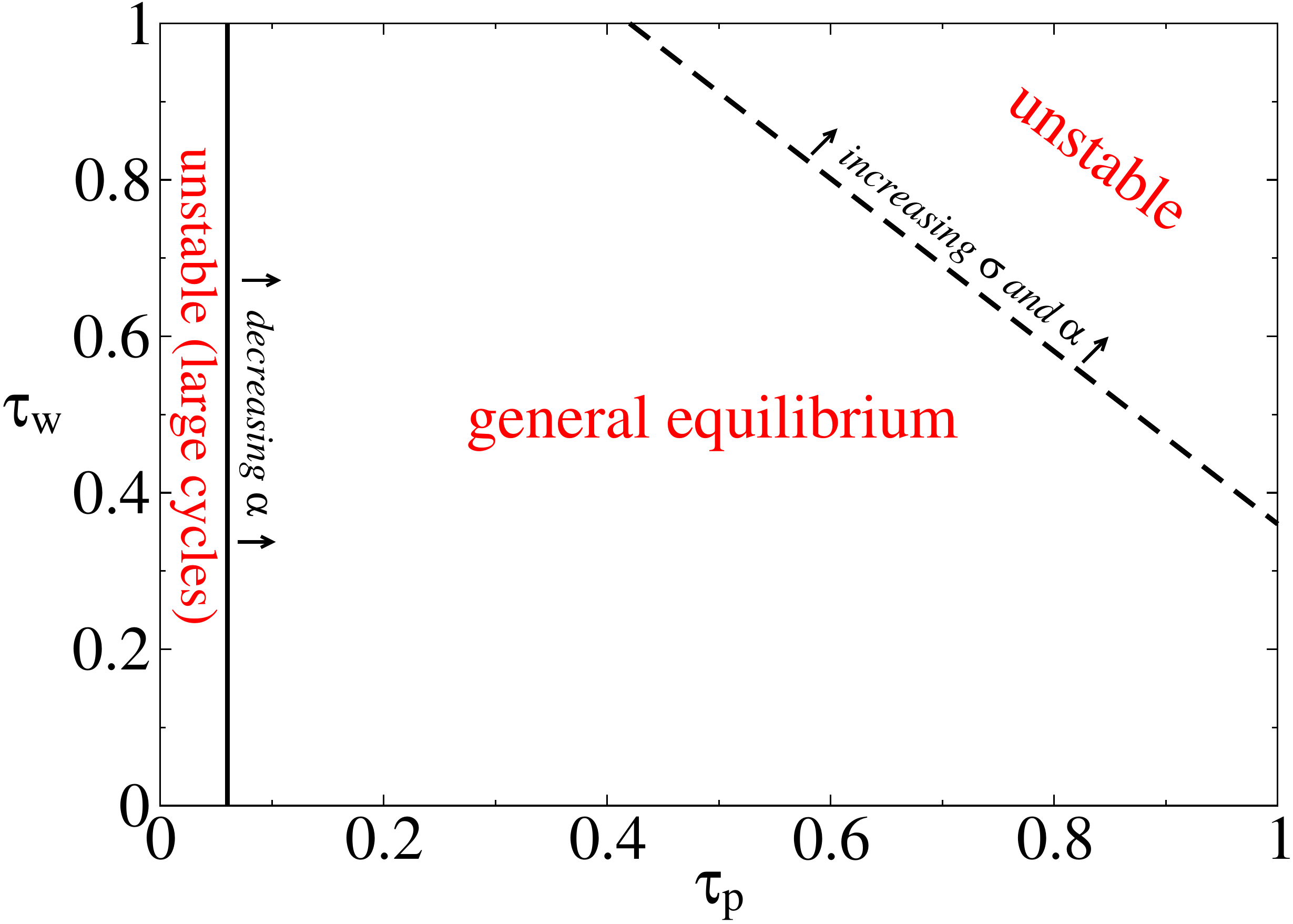}
\caption{Phase diagram of the model with $\rho_{chg}=0$ for $\theta=4/5$. Note that the unstable phase in the upper right corner disappears for values of the elasticity parameter $\theta$ sufficiently smaller than $1$ (see details in the text).}
\label{fig:fig2_pd}
\end{center}
\end{figure}

\begin{figure}[H]
\begin{center}
\includegraphics[scale=0.23]{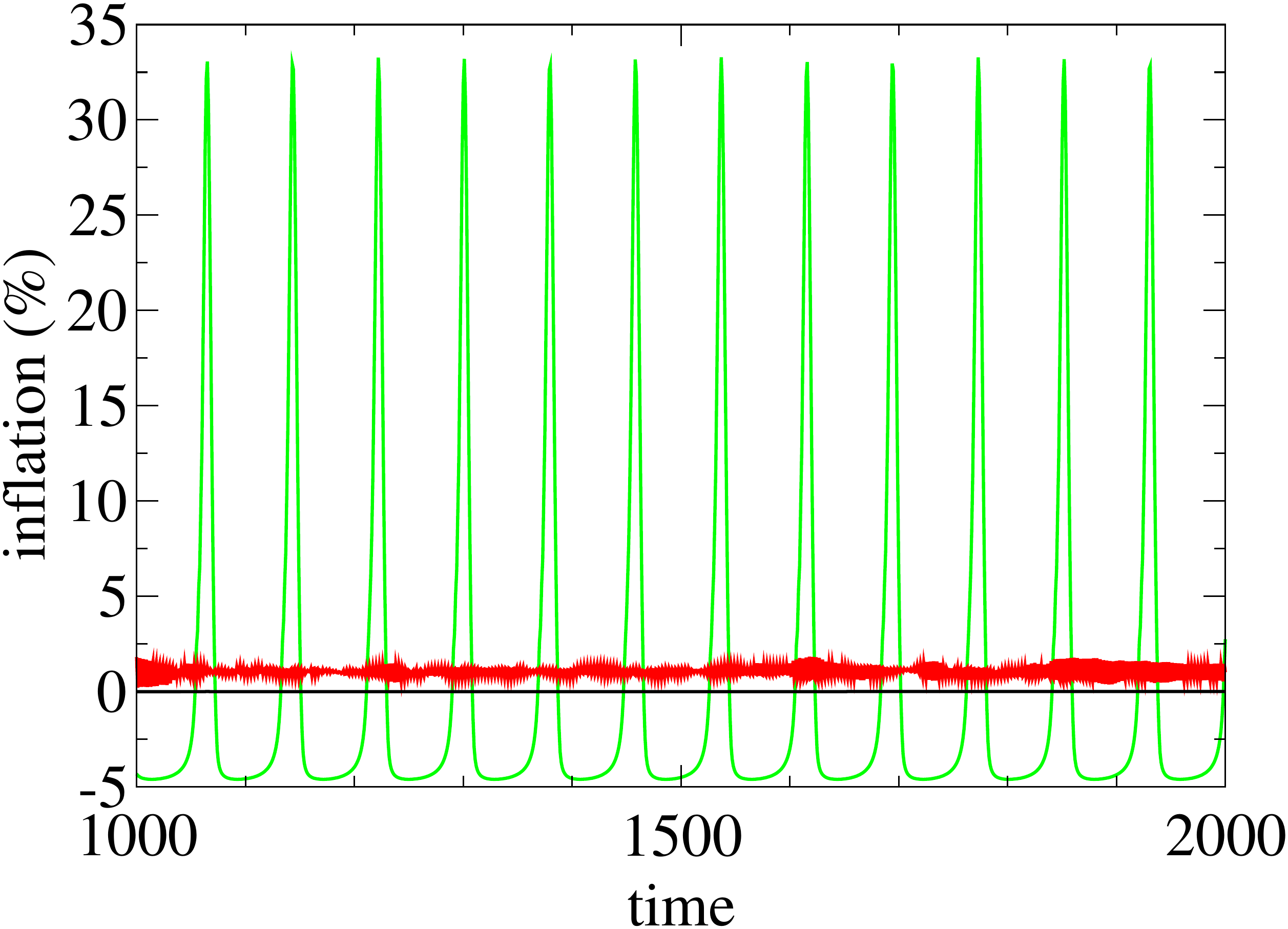}
\includegraphics[scale=0.23]{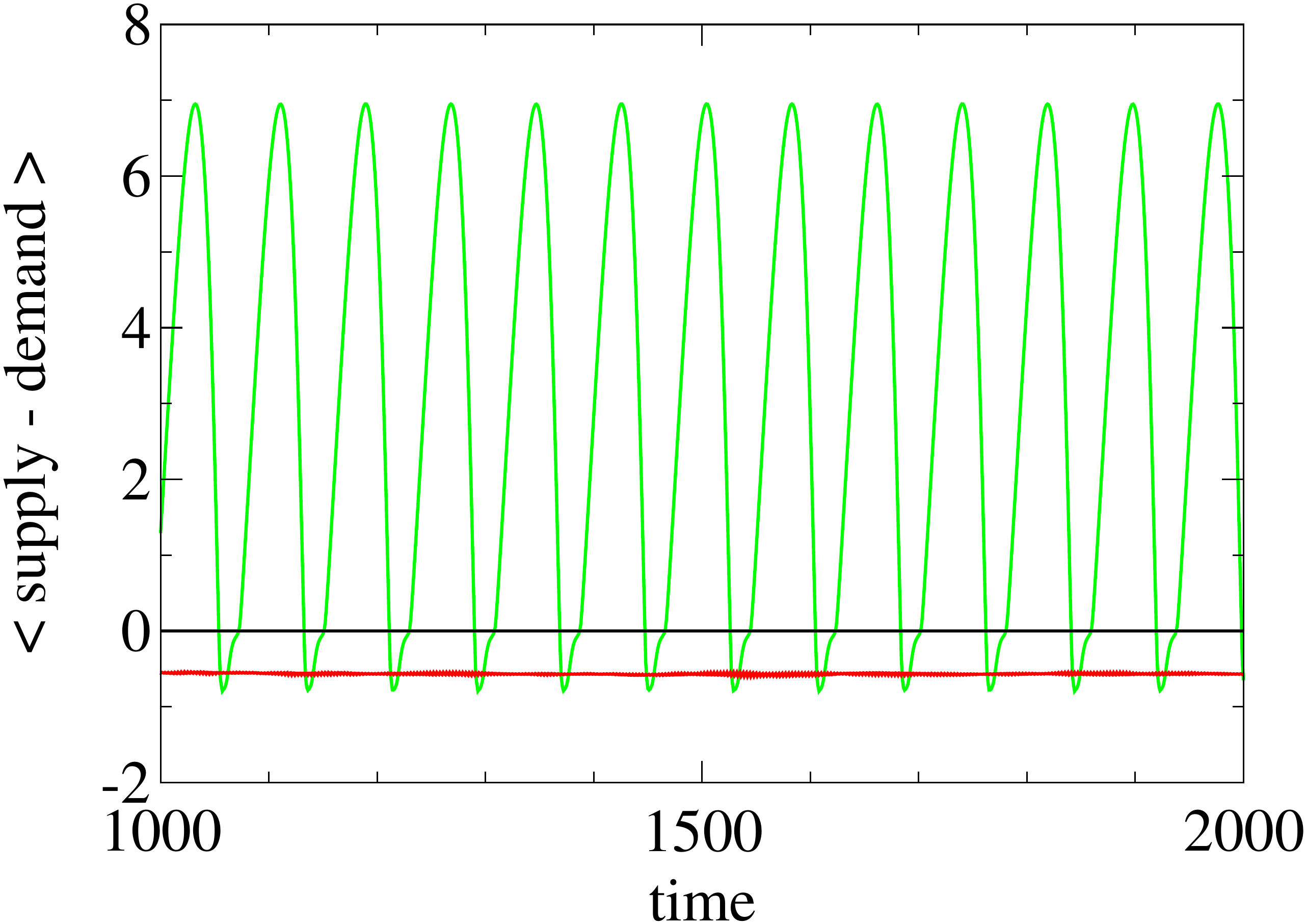}\\
\includegraphics[scale=0.23]{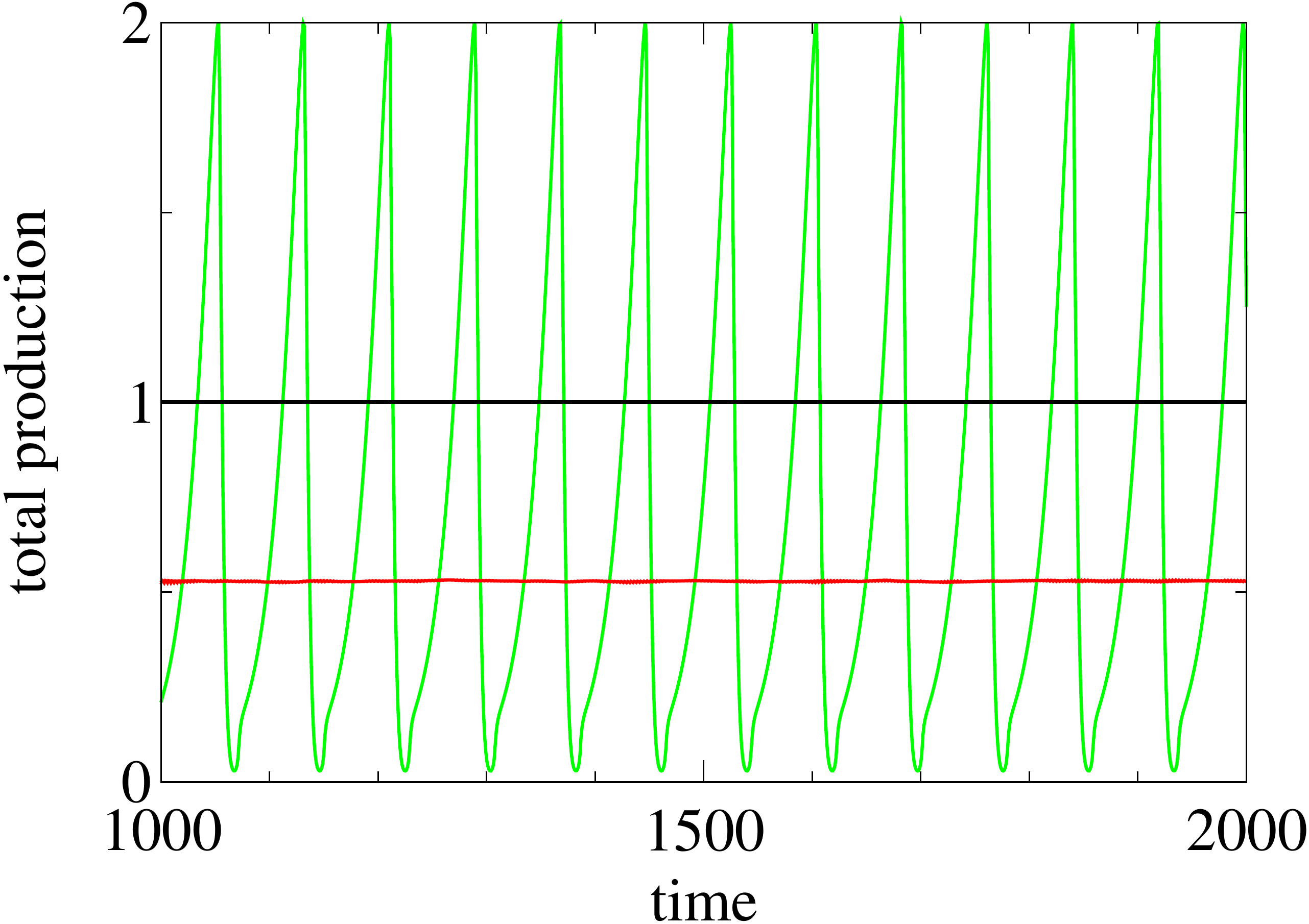}
\includegraphics[scale=0.23]{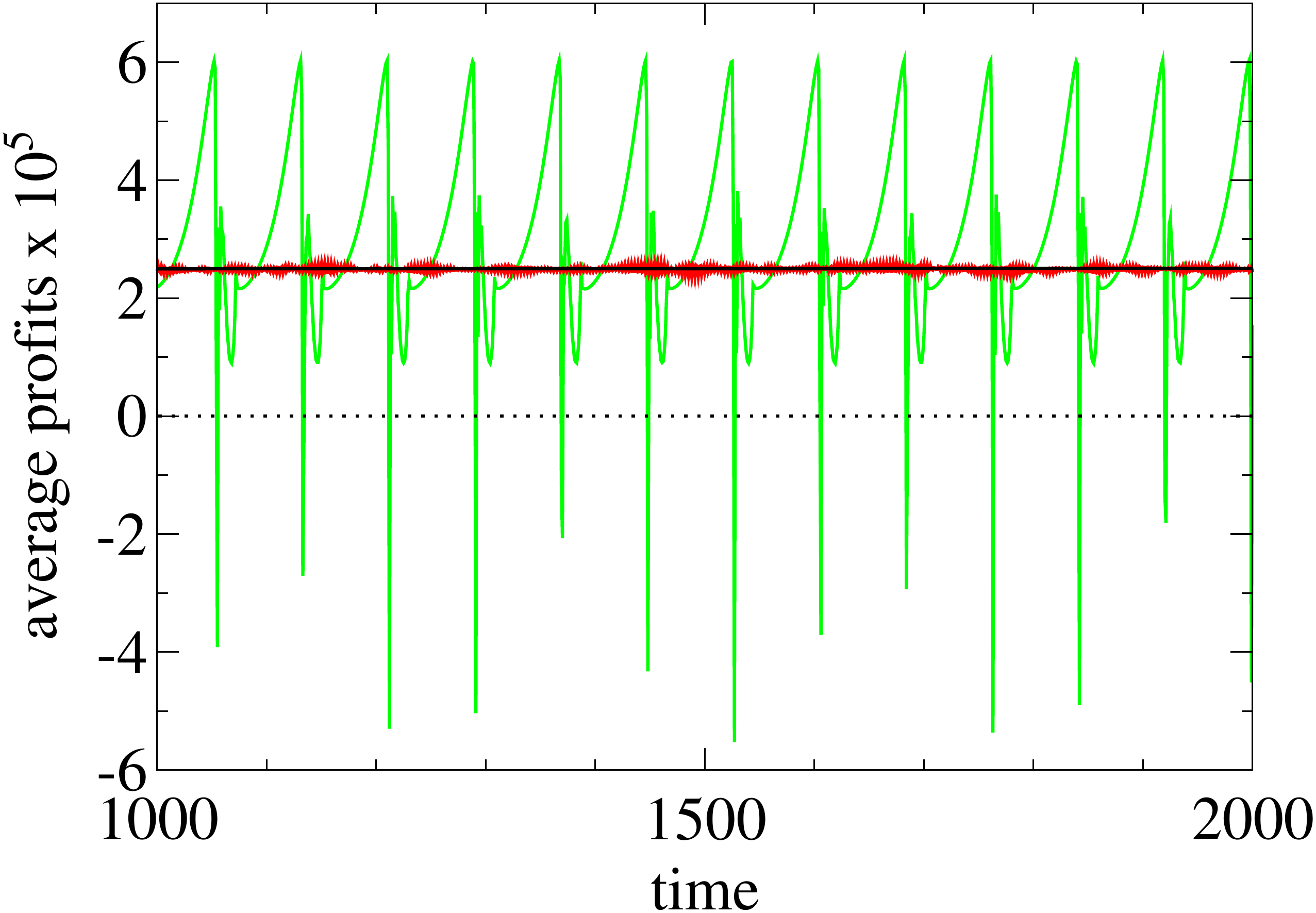}
\caption{One time-step inflation rate, average mismatch between supply and demand, total production and average firms profits as a function of time 
for the basic model ($\rho_{chg}=0$) and three different values of $\tau_p,\ \tau_w$ corresponding to different regions
of Figure~\ref{fig:fig2_pd}: $\tau_p=0.5$ and $\tau_w=0.5$ (black lines), $\tau_p=0.9$ and $\tau_w=0.9$ (red lines), $\tau_p=0.05$ and $\tau_w=0.5$ (green lines). 
Other parameters are: $\theta=4/5$, $\lambda=0.05$, $M=2000$.}
\label{fig:fig1_dyn-s125}
\end{center}
\end{figure}

\begin{figure}[h]
\begin{center}
\includegraphics[width=0.5\linewidth]{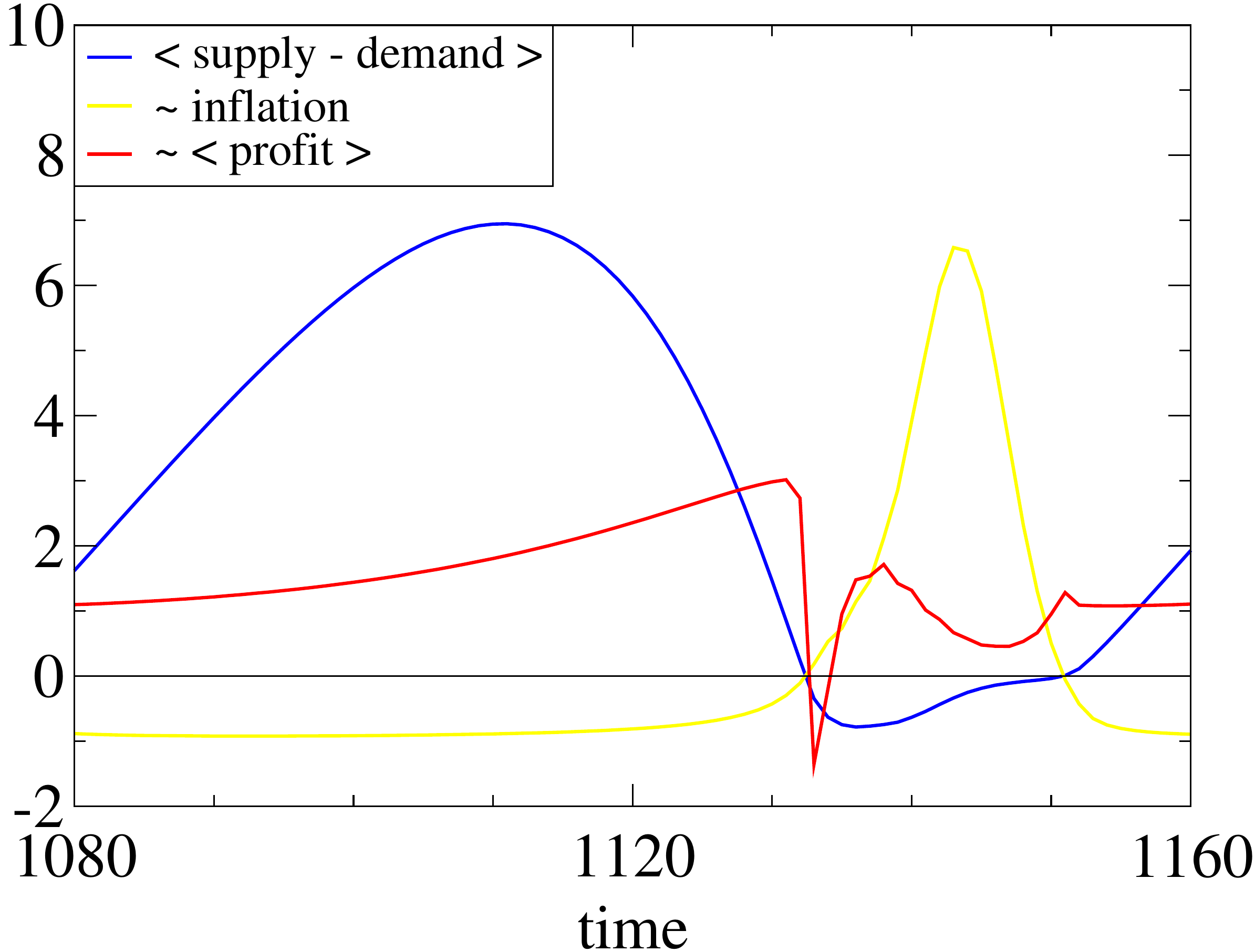}
\caption{
Average excess supply, inflation and average profit as a function of time for a cycle observed in Figure~\ref{fig:fig1_dyn-s125} for $\tau_p=0.05$ and $\tau_w=0.5$.
Inflation and profits are rescaled for illustrative purposes. Other parameters are as in Figure~\ref{fig:fig1_dyn-s125}.
}
 \label{fig:kandinsky}
\end{center}
\end{figure}

While the adjacency structure is fixed, there are only three parameters shaping the qualitative behavior of the economy in our model: the speed of price adjustment  $\tau_p,$ the speed of technological adjustment $\tau_w$ and the elasticity parameter $\theta$. In particular, for a given value of the elasticity parameter $\theta$, the speeds of price and technological adjustment determine whether the underlying general equilibrium of the economy is dynamically stable.
A numerical investigation of the model indeed reveals the presence of $3$ distinct phases of the economy (see the phase diagram depicted in Figure~\ref{fig:fig2_pd}):

\begin{enumerate}
\item For very low speed of the price adjustment process (and independently of the speed of  technological adjustment) the economy is characterized by a synchronized state with large ''cyclical volatility". As illustrated in Figure~\ref{fig:fig1_dyn-s125}, the system then oscillates between inflation and deflation, excess demand and excess 
supply, positive and negative profits. In this phase, as the prices evolve too slowly,  inventories carry the burden of adjustment.  This leads to very strong feedback effects which entail a synchronized state of the economy (as in \cite{gualdi2015tipping} and \cite{GualBouPRL}).\footnote{Note that while in~\cite{gualdi2015tipping} the main driver for the emergence of the synchronized state was the accumulation of debt, in our case the driver seems to be the accumulation / depletion of product stocks.}
Figure $\ref{fig:kandinsky}$ illustrates the processes at play during a cycle. During a first phase, production is larger than demand and firms build up stocks while the price adjusts downwards. During a second phase (after the inflection of the supply-demand curve), demand is larger than supply,  the stocks get built down but prices keep adjusting downwards (there is still excess supply as the inventory is non-empty). The price keeps adjusting downwards until the stocks are completely depleted. At this stage, excess demand is at a maximum and profits at a minimum because prices are low and stocks are empty. Yet, in absence of buffer stocks, prices become  more volatile and increase rapidly until excess demand is absorbed and a new cycle starts.
\item For intermediate speeds of price adjustment the system converges to equilibrium (see Figure~\ref{fig:fig1_dyn-s125}). This phase is largely predominant and will be the setting used for the analysis of distributional firms and network properties carried out in the following.
\item When both price and technological adjustment speeds are high the economy reaches an ``excess demand'' phase with rationing. There is persistent mismatch between supply and demand, positive inflation and sustained volatility in the network (see Figure~\ref{fig:fig1_dyn-s125} as well).\footnote{Note that disequilibrium appears to materialize via inflation and excess demand (the inflation is small but persistently positive in the upper-left corner of  Figure~\ref{fig:fig1_dyn-s125}) rather than via deflation and excess supply. This is however an aggregate view which averages excess demand and excess supply, increasing and decreasing prices. Yet, there is a bias towards inflation and excess demand because the price adjustment process (Equation $\ref{pricefric}$) induces an asymmetry in the magnitude of the price adjustments upwards and downwards.} 
Note however that this phase disappears for values of the elasticity parameter $\theta$ below a threshold $\theta^*$ ($\theta^*\sim 5/9 $). 
In practice, this phase is less robust and manifests itself only for values of $\theta$ close enough to $1$.
\end{enumerate}
 
Hence,  despite its simplicity (there is no evolution of the adjacency structure of the network at this stage), the model displays a very rich taxonomy of behavior: general equilibrium, rationing and periodic ``overproduction" crisis.

 The key determinant of the transition between the stable general equilibrium phase and the unstable ``excess demand" phase is the relative speed of price ($\tau_p$) and technological ($\tau_w$)  adjustment. The faster the relative speed of price adjustment, the more stable is the system. Yet, 
 the stability range increases as the absolute speed of price adjustment decreases
\footnote{These  results are reminiscent of those obtained in \cite{bonart2014}: the larger 
the intrinsic volatility of the system (in our setting, the higher the elasticity of 
substitution), the slower the adjustment processes shall be for 
the system to be stable.}.   
 Moreover, the size of the stable region increases both as the elasticity of substitution decreases and as the share of labor in the input mix increases (i.e. for larger values of the parameter $\alpha$ in the production function).
% There exists a critical value $\theta^*$ ($\theta^*\sim 5/9 $ for the parameter setting used in figure  \ref{fig:fig2_pd}) such that for $\theta\leq \theta^*,$ the unstable region disappears and the system converges to equilibrium independently of the speeds of price and technological adjustment.

\subsubsection{Network Structure}

In order to investigate numerically the asymptotic properties of the production network, we have performed a second series of simulations in which the adjacency structure evolves  (i.e. $\rho_{chg}>0$) according to the competitiveness of firms (see $\ref{netdyn}$). The results we  observe are  consistent with Proposition $\ref{propme}$ and  robust. As long as the parameters are in the predominant general-equilibrium phase, the distribution of firms' in-degree converges to a power-law with an exponent close to $1$ (i.e. the Zipf distribution) for large enough in-degrees. Convergence occurs independently of the initial distribution of firms' sizes and both the power-law nature and the exponent of the limit distribution do not depend on the parameters used in the simulations, in particular  the elasticity of substitution. 

In more details, we observe that the convergence to the invariant distribution is extremely long (approximately $10^6$ time steps) but after convergence the results of a particular realization are representative of the average across multiple realizations.  From a qualitative point of view, we observe two different regimes:

\begin{itemize}
\item In the predominant region of the parameter space, which corresponds to the general equilibrium phase described in the preceding section, we observe convergence to an invariant distribution characterized by strong heterogeneity of firms in-degrees (provided there is initially a minimal amount of heterogeneity\footnote{That is unless all the firms are initialized with the same number of suppliers.}). In order to characterize the tail of the distribution, we follow the procedure put forward in \cite{clauset2009} and perform Kolmogorov-Smirnov test first on a series of $100$ simulations corresponding to different seeds and initial networks for fixed parameters.
We obtain a very good fit to  a power-law with exponent close to $1$ (i.e. the Zipf distribution). Figure ~\ref{fig:fig4_incoming} illustrates the result for a given realization of the model (after $4\ 10^6$ time steps) while  Table~\ref{pltable} provides statistical evidence (over $100$ simulations) about the power-law nature of the tail of the in-degree distribution, as well as that of the size of firms (measured either by their sales or by the total incoming weight they receive).
We have then performed extensive simulations both over different pairs $\tau_p,\ \tau_w$ in the general equilibrium phase (20 simulations) and over different pairs $p_{new}, \ \rho_{chg} \in[0.01,0.2]$ (50 simulations). In both cases we do not observe departures from the results shown in Table~\ref{pltable}. We have also checked different values of the elasticity of substitution as well as different mark-ups $\lambda$, which also give results perfectly consistent with Table~\ref{pltable} confirming that both the power-law nature of the tail and its exponent are stable features of the model. Note however that, since the exponent is independent of the parameters, the distribution cannot be scale-free on the whole  range because the average in-degree is fixed by the total number of links in the network.

\begin{remark} For negative values of $\theta$ (corresponding to complementary goods) we still observe convergence to a broad distribution, but in this case the power law regime is recovered only for intermediate in-degrees. Indeed, the tail of the distribution corresponds in this setting to a pathological case where  firms' competitiveness is bounded by a minimum number of suppliers (equal to one). From this point of view the model presented in this paper has to be generalized (for example by allowing heterogenous pre-factors in the production function) in order to be consistent with the presence of complementary goods.
\end{remark}

\begin{remark}  We do not account for inter-temporal substitution of consumption in the dynamics considered above. Indeed, in absence of investment opportunities for firms, there cannot be a demand for credit from the productive side and  savings wouldn't bear an interest rate. Therefore, the only motive for savings would be to hedge against potential income shocks. Such savings could potentially smoothen the macro-economic fluctuations and hence decrease the size of the unstable phase of the economy (see figure \ref{fig:fig2_pd}).  However, they are unlikely to have any impact on the structure of the production networks that emerge within the stable phase.    

\end{remark}

\begin{table}[h]
\center
\begin{tabular}{|c|c|c|c|}
\hline
\textbf{variable} & \textbf{exponent} & \textbf{$x_{min}$} & \textbf{KS p-value}\\
\hline 
in-degree & $1.07\pm 0.01$ & $13\pm6$ & $0.69\pm0.21$ \\
\hline 
 total incoming weight & $1.26\pm0.06$ & $1.25\pm0.32$ & $0.80\pm0.23$  \\
 \hline
 total sales & $1.19\pm0.06$ & $2\pm 0.4 \ 10^{-3}$ & $0.57\pm0.22$ \\
 \hline
 \end{tabular}
 \caption{Results of statistical tests for power-law tails of the distributions plotted in Figure~\ref{fig:fig4_incoming}. This results are averages over $100$ realizations with different seeds and initial networks (but fixed parameters corresponding to Fig.~\ref{fig:fig4_incoming}). The estimated exponent refers to the complementary cumulative distribution. All computations are performed using the R package ``PoweRlaw". }
 \label{pltable}
\end{table}

 \begin{figure}[h!]
\begin{center}
\includegraphics[width=0.4\linewidth]{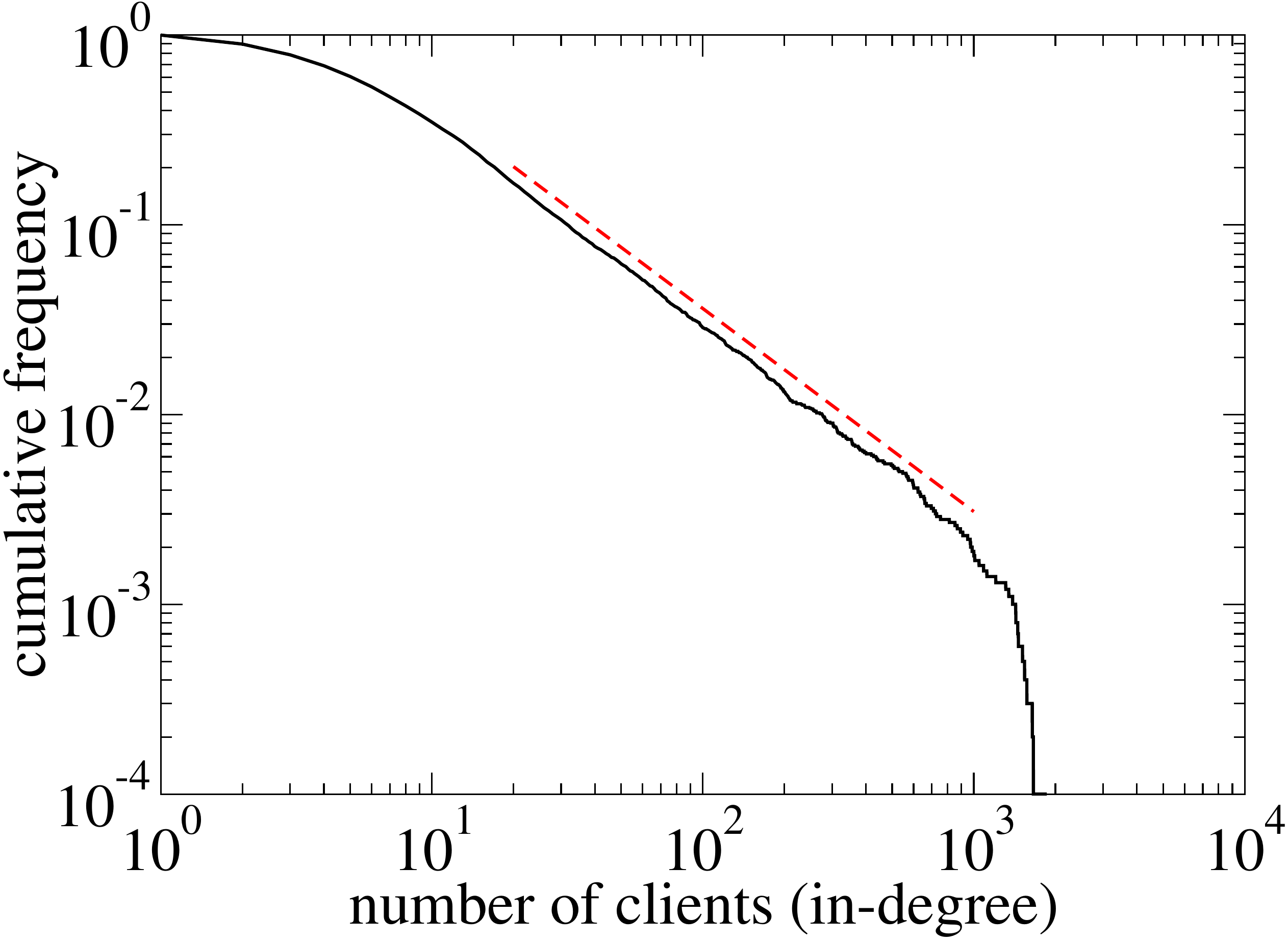}
\includegraphics[width=0.4\linewidth]{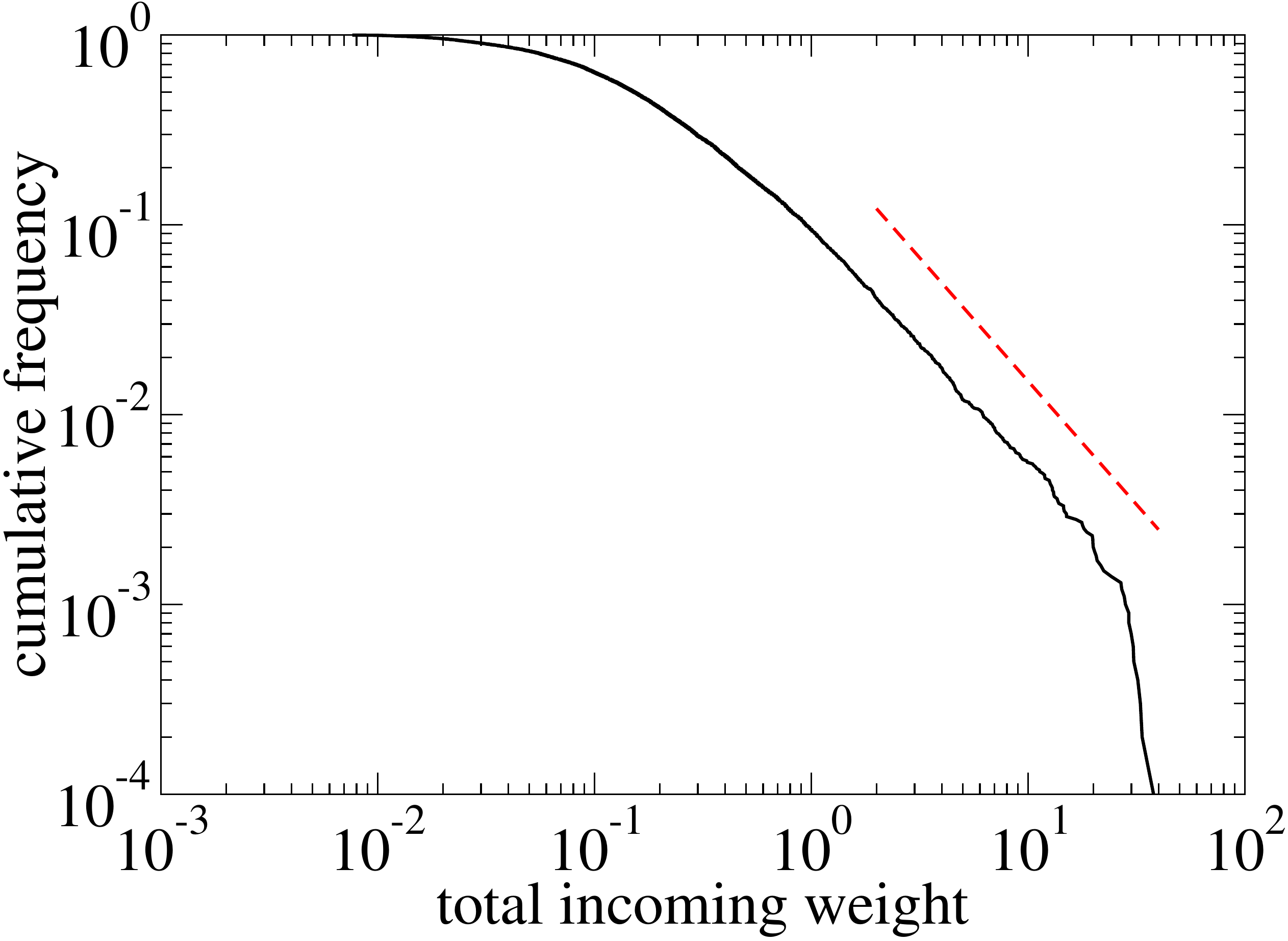}\\
\includegraphics[width=0.4\linewidth]{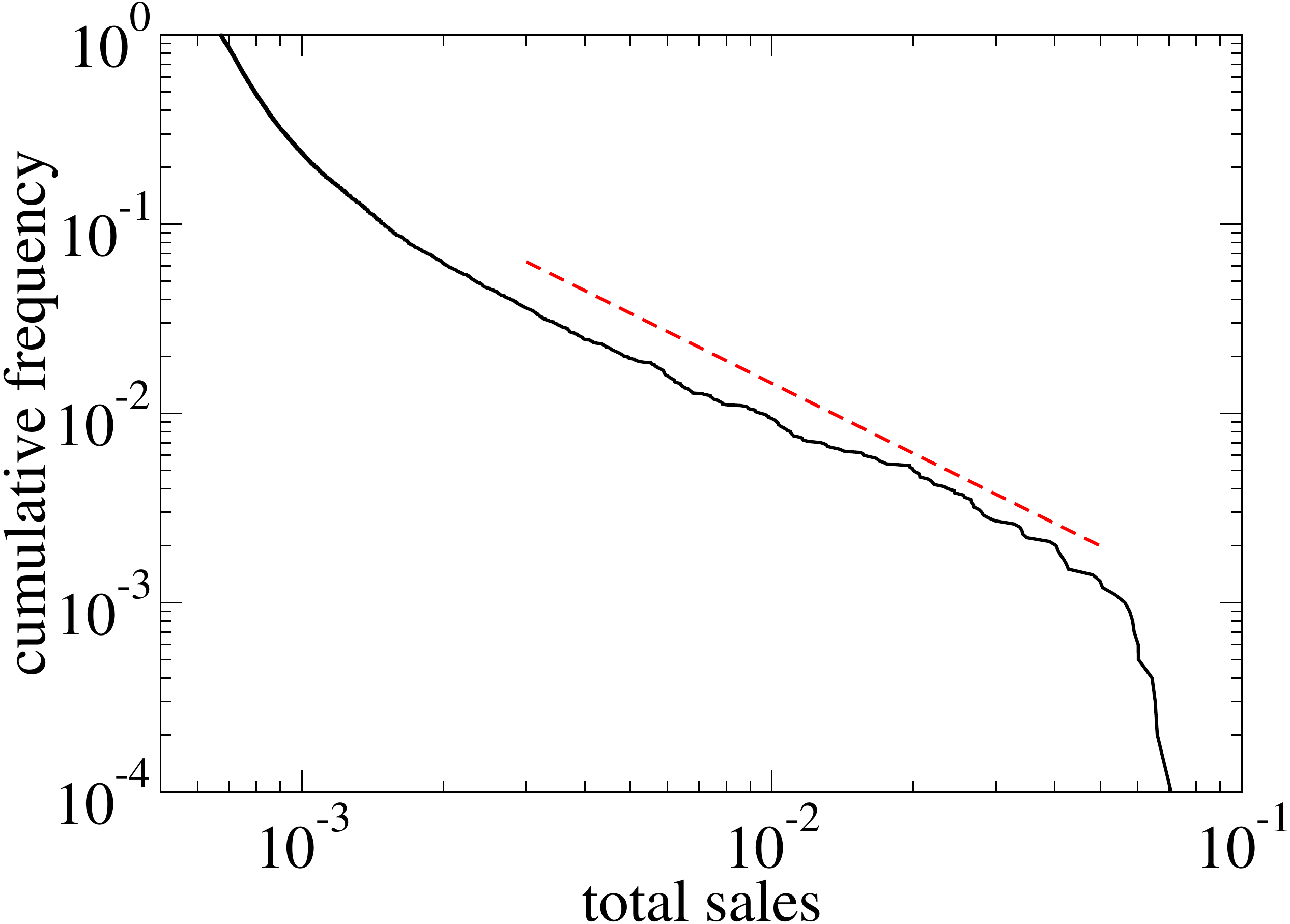}
\includegraphics[width=0.4\linewidth]{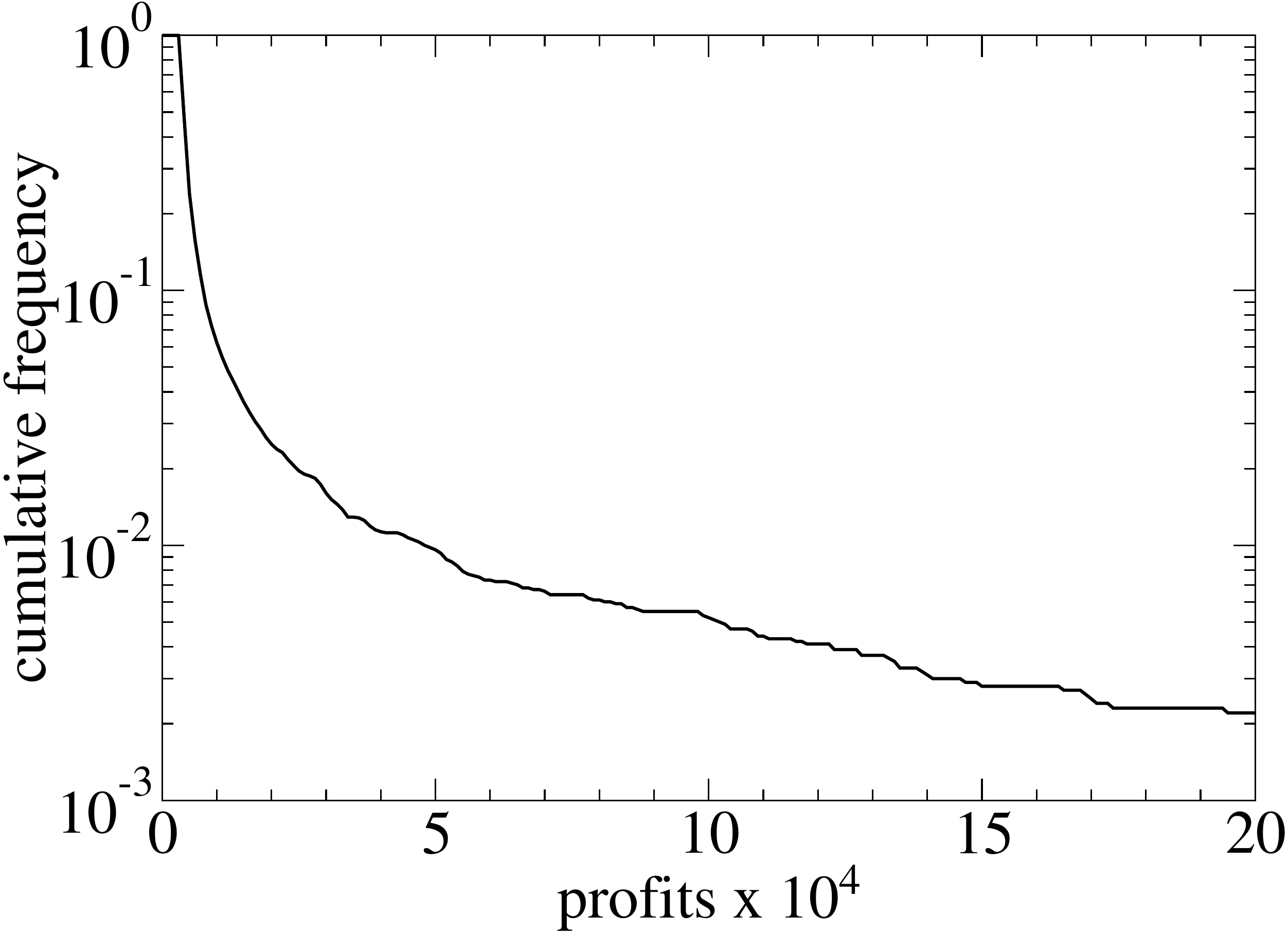}
\caption{
Basic firms statistics after $4\ 10^6$ time steps for a particular run of the model with 
$\rho_{chg}=0.01$ and $\rho_{new}=0.05$.
\emph{Top:} Cumulative frequency for the number of incoming links (clients) and 
total incoming weight (sum over all clients weights). 
\emph{Bottom:} Cumulative frequency for total sales and profits histogram.
Dashed red lines correspond to the power-law fit performed with the R package poweRlaw on the 
distributions.  Average results of the Kolmogorov-Smirnov test and fitted parameters are summarized in Table~\ref{pltable}.
Other parameters are: $\theta=0.5$, $\tau_p=\tau_w=0.8$ and $M=10000$.}
\label{fig:fig4_incoming}
\end{center}
\end{figure}

\item  If the parameters of the model lie outside the general equilibrium phase depicted in Figure~\ref{fig:fig2_pd} (or if firms  all have the same number of suppliers), firms' characteristics remain relatively homogeneous: we do not observe the emergence of fat-tails and the support of the asymptotic distribution is only mildly larger than at initialization. Proposition~\ref{propme} cannot be applied either because prices fluctuate wildly due to network synchronization effects (hence there is no clear correspondence between the price of a firm and its competitiveness) or because  firms do not differentiate in terms of competitiveness (i.e. if the number of suppliers is the same for all firms). 

\item Finally, in the extreme case where there is no entry (i.e. $p_{new}=0$) the less competitive firms definitively exit the market and only approximately $1\%$ of the initial number of firms survive. The network converges to a steady-state with a hierarchical structure consistent with Proposition~\ref{propns} (see also the discussion following Remark~\ref{remark3}). Figure~\ref{fig:no_entry} illustrates this situation via  a scatter plot of the final number of clients as a function of the number of suppliers. Different in-degrees for the same out-degree are due to the fact that depending on the particular place in the supply chain firms with the same production capabilities may offer slightly different prices.

 \begin{figure}[h]
\begin{center}
\includegraphics[width=\linewidth]{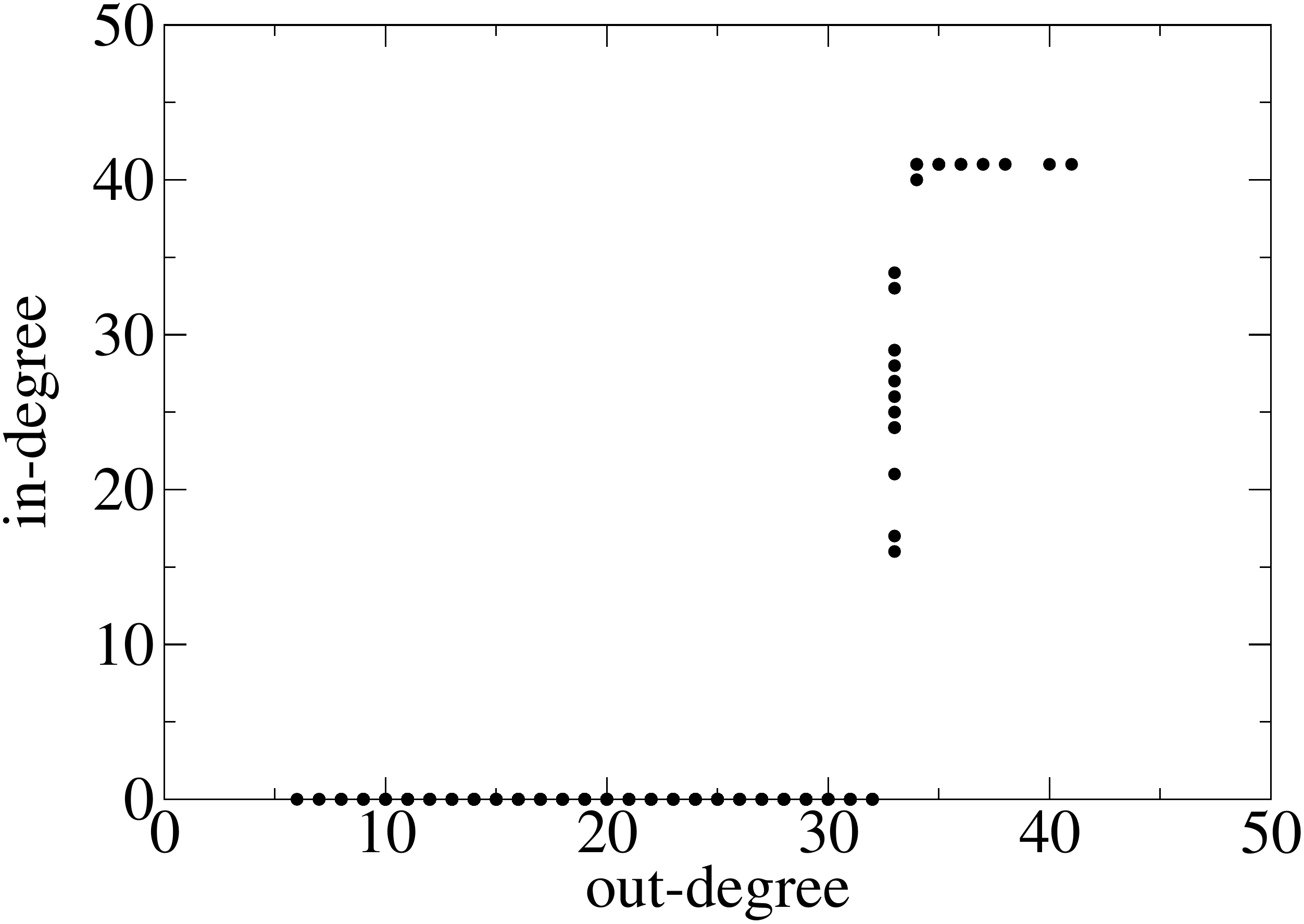}
\caption{
Illustration of the extreme situation where there is no entry ($p_{new}=0$). At the steady state only more competitive firms survive ($\sim 1\%$ of the total).
In this plot we show the final in-degree as a function of the number of suppliers (out-degree). There is a direct relation between the out-degree and the in-degree
consistently with Proposition~\ref{propns}. Firms with the same number of suppliers only differentiate for their
position in the supply chain. Parameters are the same as in Figure~\ref{fig:fig4_incoming}.}
\label{fig:no_entry}
\end{center}
\end{figure}

\end{itemize}

\begin{remark} The results of this subsection offer a partial reciprocal to Remark $\ref{redux}:$ it is necessary that the equivalence with the simplified model sketched in remark $\ref{redux}$ holds, i.e. that convergence to general equilibrium occurs when the network is fixed, for the model to display a consistent asymptotic behavior.
\end{remark}

To sum up, under very general conditions (as long as $\tau_p$ and $\tau_w$ lie in the general equilibrium phase), our model converges to a self-organized state characterized by heterogenous firm and network properties. The distributional properties appear as independent of the parameters of the simulations and match stylized empirical facts about the  
distribution of firms' size and connectivity. The introduction of a minimal amount of technological change in a simple model of monopolistic competition suffices to  reproduce  key stylized facts about 
the distribution of firms' size and the structure of production networks. Indeed, it suffices to allow firms to shift between suppliers rather than to consider that the production infrastructure is fixed to observe the endogenous emergence of scale-free structures.  This result is in line with the literature \cite[e.g.][]{hopenhayn1992entry,rossi2007establishment,luttmer2007selection} but very parsimonious in terms of rationality and complexity. Moreover, beyond size and degree distributions, the model is able to reproduce key stylized facts of firms' demographics, as described below.

\subsection{Distribution of firms' growth rates}

A first major stylized fact of firms' demographics is  that the growth rates of 
firms are distributed according to a ``tent-shaped" double-exponential 
distribution \cite[see][]{bottazzi2006}. As illustrated in 
Figure~\ref{fig:fig3_growth}, our model reproduces well this type of Laplace 
distributions. The exponential distribution is a good fit for the central part of the support,
while extreme values of the growth rates exhibit fatter tails.
We show results for two different time intervals used in the computation of growth rates,
in both cases we take the last (up to a maximum of $30$) recordings of firms sales.
As one can see shorter time intervals naturally result in a steeper descent of the exponential.
To provide an estimate of the parameter of the distribution we fit the two distributions via a maximum-likelihood fit with
$f(x)=b_\pm\exp{(-b_\pm|x|)}$ and find: $b_+ = 7.11\pm 0.03$ (red dashed line)  and $b_+ = 18.05\pm 0.06$ (black dashed line)
for $x>0$; $b_- = 8.95 \pm 0.04 $ (red dashed line) and $b_- = 19.22 \pm 0.06$ (black dashed line) for $x<0$.
All fits are done in the interval $|x|\in [0.05,0.5]$ and are just for illustrative purposes since the exponential distribution 
cannot fit data in all the range failing a statistical test. 
As for the results in the previous subsection we do not observe any dependence on the parameters of the model with the exception in this
case of the $\rho_{chg}$ parameter. Higher values of $\rho_{chg}$ indeed result in a more gradual descent of the exponential. This is a rather
intuitive result since increasing $\rho_{chg}$ results in more link swaps per time step and therefore allows firms to grow/decline faster.

Following \cite{arthur1994}, \citet{bottazzi2006} put forward the 
fact that a Laplace type of distribution emerges because market success is 
cumulative or self-reinforcing. In their ``island-based" model, this 
self-reinforcing process is hard-wired into the model:   
 \emph{``we model this idea using a process whereby the probability for a given 
firm to obtain new opportunities depends on the number of opportunities already 
caught."} In our setting, ``self-reinforcing success" is also at play but it 
emerges endogenously. Indeed, the price-setting process (see equation 
$\ref{pricefric}$) is such that whenever a firm gains a new consumer, its price 
increases (directly but also indirectly through the increased demand that it 
addresses to his own suppliers) and hence its competitiveness decreases. However, 
the larger the firm is, the weaker the effect of an additional consumer is on its 
price and hence the more competitive it remains. Therefore, larger firms are 
more competitive and  can seize more frequently new business opportunities. 

.

\subsection{Size and variance of growth rates}

\begin{figure}[h!]
\begin{center}
\includegraphics[width=0.45\linewidth]{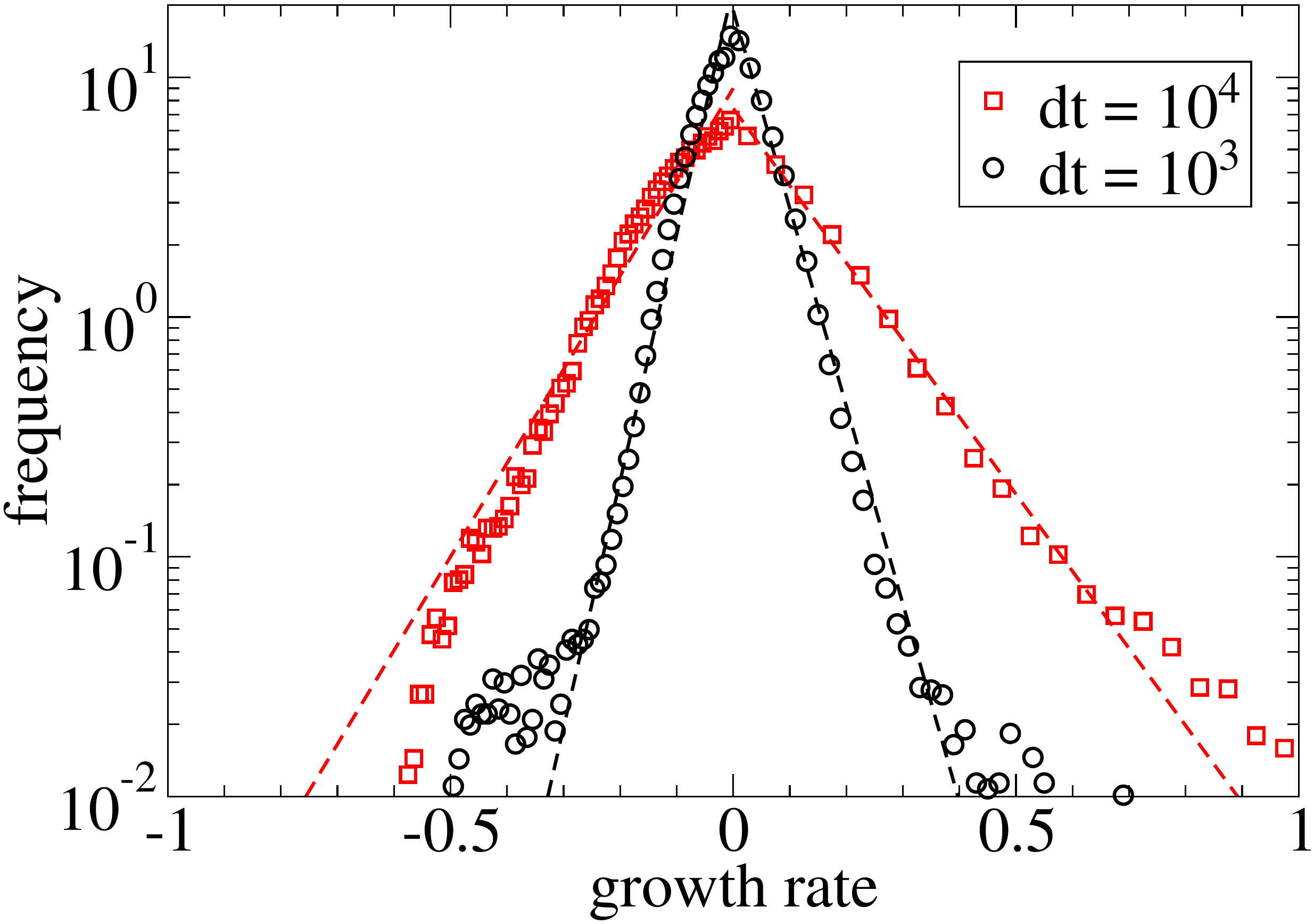}\quad
\includegraphics[width=0.45\linewidth]{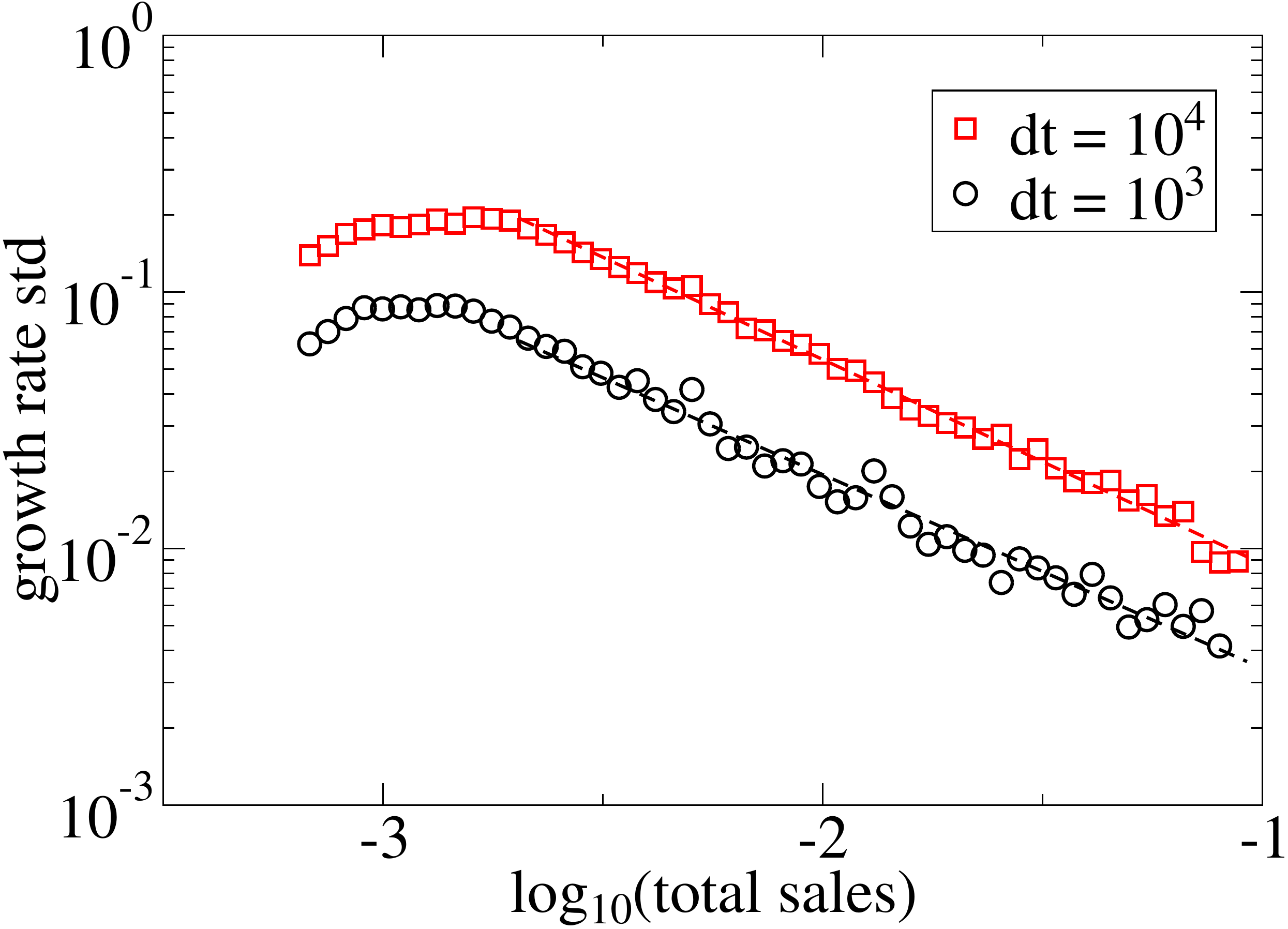}
\caption{Distribution of firms' growth rates (left plot) and standard deviation of firms growth rates as a function of total sales (right)
after $4\ 10^6$ time steps for the same parameter setting
as in Figure~\ref{fig:fig4_incoming}.
Different symbols / colors correspond
to different time intervals to compute growth rates (for each firm we plot only the last $30$ rates).
}
 \label{fig:fig3_growth}
\end{center}
\end{figure}

A second major stylized fact of firms' demographics concerns the negative relation between the variance of growth rates and the 
 size of firms \cite[see][and references therein]{coad2009growth}. More precisely, the empirical literature \cite[see e.g.][]{amaral} documents a relation of the form $\sigma(s)\sim s^{-\beta},$ where $s$ is the size of the firm and $\sigma(s)$ the variance of growth rates for firms of size $s.$   In order to analyze the fit of the model to this relation, we follow \citet{amaral} and  group firms according to their sales in log-bins with base = $1.1$ and compute the standard deviation of growth rates within each bin. As one can see in Figure~\ref{fig:fig3_growth}, numerical data perfectly matches the functional relation
found in the empirical literature. A linear fit of the logarithm of the standard deviation versus the logarithm of sales gives 
values of $\beta$ equal to $0.075\pm 0.001$ (for $dt=10^4$) and $0.072 \pm 0.002$ (for $dt=10^3$), both with $0.1\%$ significance.
In this respect, our estimate of the parameter $\beta$ is smaller by a factor of $3$ than the one found in \citep{amaral} and \citep{bottazzi2001}.
We do not observe any dependance of the estimate with respect to parameters of the model.

Finally, we also check departures from Gibrat's Law by regressing the model
\begin{equation}
\log{(s_t/\bar{s}_t)} = \gamma \log{(s_{t-1}/\bar{s}_{t-1})} + \epsilon
\end{equation}
where $s_t$ is the size of the firm at time $t$ and $\bar{s}_t$ is the average over all firms.
This is a standard test in the empirical literature \cite[see][and references therein]{bottazzi2001} which highlights deviations
from the ''law of proportionate effect'' (where the proportional rate of growth of a firm is independent of its size) as long as the value of $\gamma$ is found
significantly different than $1$. In our case we always find values of $\gamma$ ranging from $0.993$ for $dt=10^3$ to $0.97$ for $dt=10^4$ (both with a negligible error
and $0.1\%$ significance) implying a mild reversion to the mean effect, i.e. smaller firms tend to grow faster, at least on longer time scales.

\subsection{Exit rates and age}
 
\begin{figure}[h!]
\begin{center}
\includegraphics[width=0.45\linewidth]{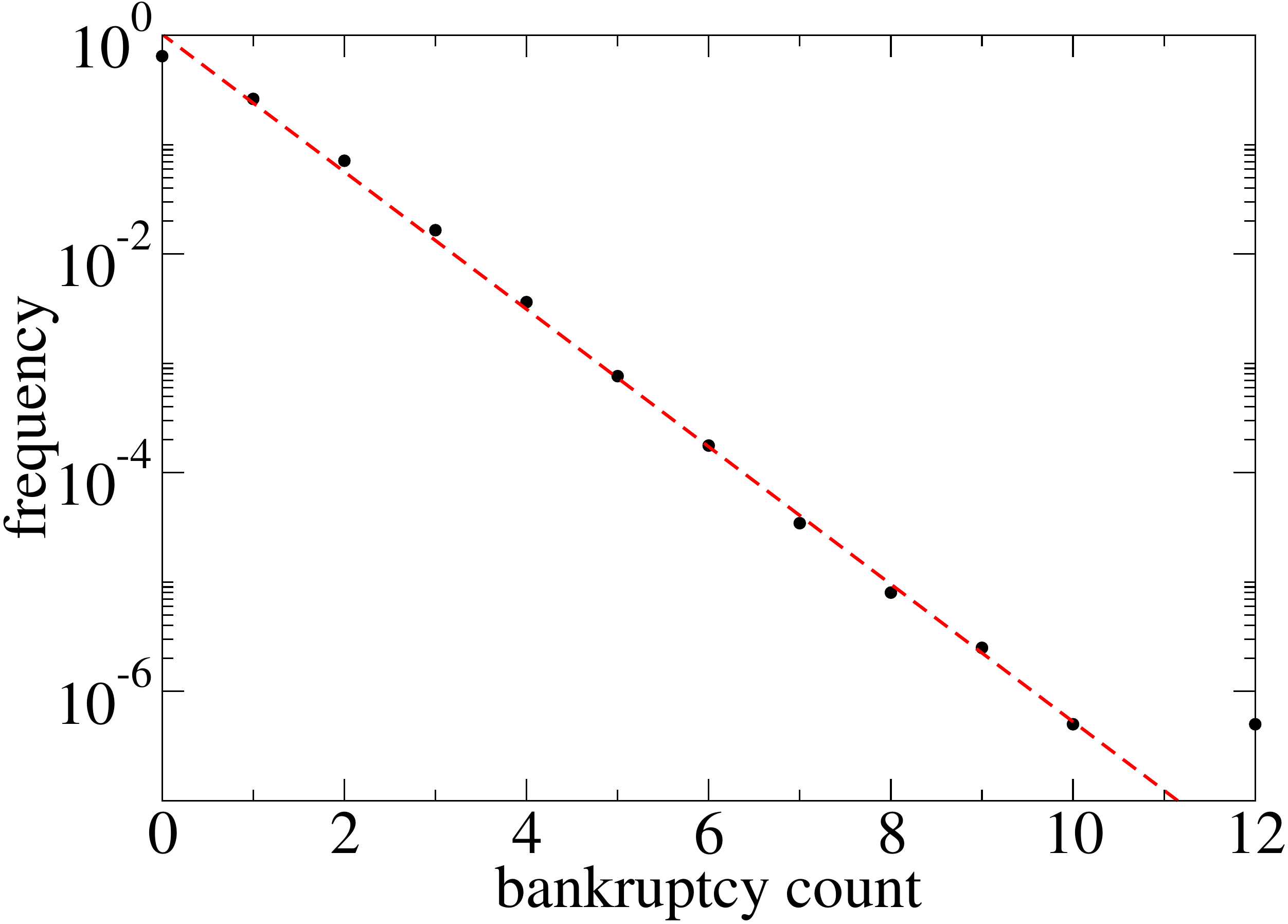}\quad\quad
\includegraphics[width=0.45\linewidth]{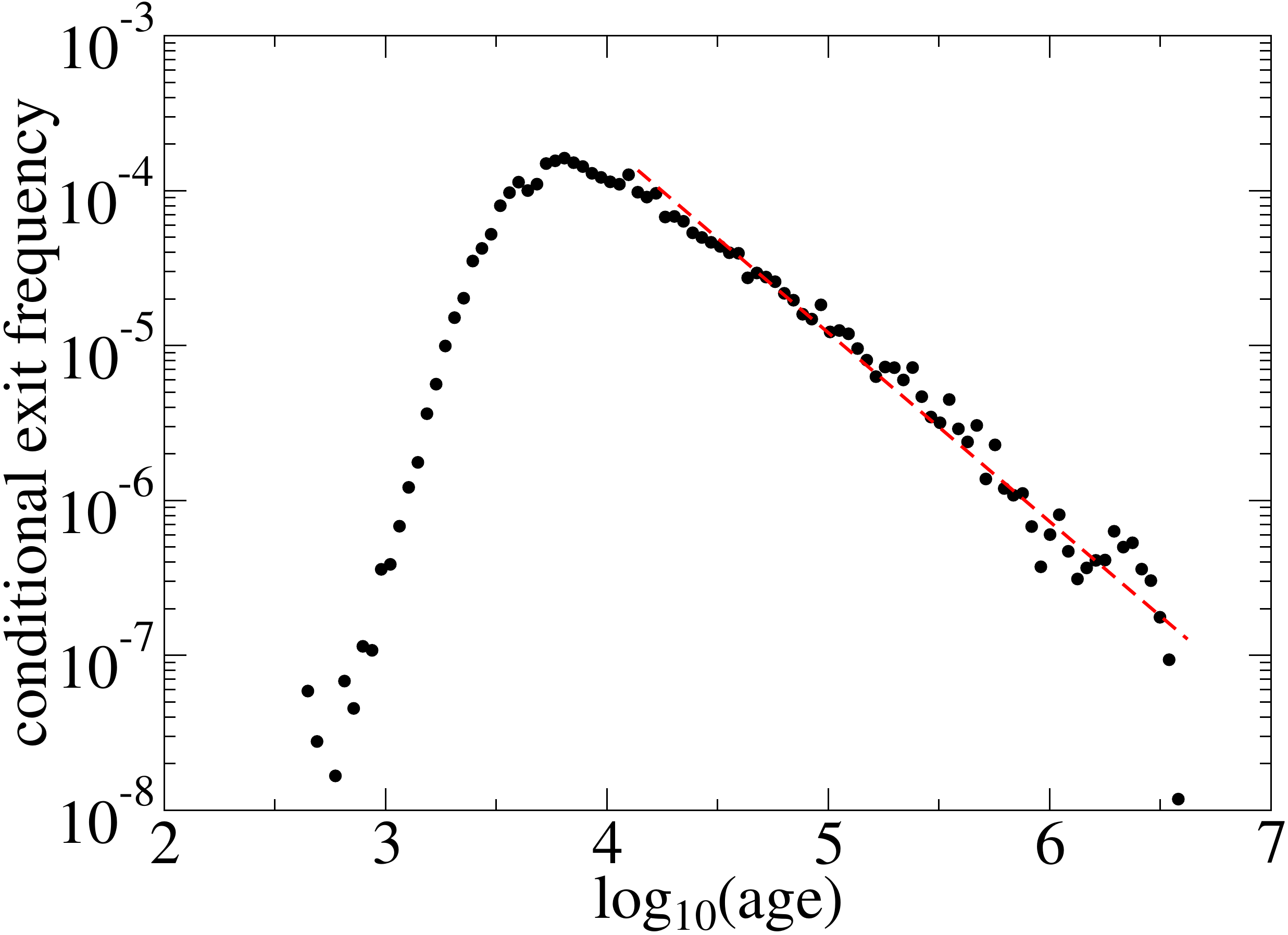}
\caption{Distribution of the number of bankruptcies per unit of time (left plot) and distribution of exit frequencies
as a function of the age of the firm (right plot).
}
 \label{fig:fig4_exit}
\end{center}
\end{figure}

 We conclude this empirical section with results about firms' exit rates.
 In Figure~\ref{fig:fig4_exit} (left) we plot the distribution of firm exit rates, i.e. the
 number of bankrupted firms per time step, which is very well approximated by an
 exponential of parameter $\nicefrac{1}{2}.$
 
 The empirical literature about firms' exit rates highlights that there is a negative correlation
 between firm exit rates and both size and age \cite[see e.g.][]{klette2004innovating}. 
 While in our model we cannot study the relation  between exit and size (our firms exit when they are of size $0$ by definition,) we can measure
 the frequency of exits conditional on firm age. 
 In order to do so, we record in the last $2\ 10^6$ time steps of the simulation both the time of each bankruptcy and the age of the bankrupted firm. 
 We then bin firms' ages in log-bins with base $1.1$ and compute the average 
 frequency of bankruptcies per unit of time in any bin. Finally, we compute the ratio between the
 bankruptcy frequency and the stationary frequency of active firms in any bin.
As one can see in Figure~\ref{fig:fig4_exit} (right) after a typical time of order $\bar{k}/\rho_{chg}$, older firms have an exponentially decaying probability of exiting.\footnote{This is because at each time step $\rho_{chg}$ firms change one supplier. If the average degree in the network is $\bar{k},$ the expected waiting time for a firm to exit is $\bar{k}/\rho_{chg}.$} An exponential fit $\sim \exp{(-a t)}$ of the tail of the distribution is a good approximation and gives $a = 0.12$. 

Note that the scope of this section is just to give a qualitative understanding of the relation between age and exit probability. Indeed, the model cannot yield a realistic age distribution of firms because age and size are too strongly connected in our framework: the random selection of new suppliers implies that large firms are there since longer times. Hence, the age distribution of firms reflects the size distribution: it is characterized by power-law tails. This is clearly an unrealistic feature of the model and it also alters the relationship between age and exit rate. 
 
\section{Conclusion}
  
The model of monopolistic competition on the markets for intermediate goods is central both to the international trade and the endogenous growth
literature  \cite[][]{ethier1982,romer1990}.  In this paper, we develop a simple agent-based dynamic extension of this model in order to investigate the endogenous formation of production networks. 
The model subsumes the standard general equilibrium approach and robustly reproduces key stylized facts of firms' demographics. First and foremost, the model shows that competition between intermediate good producers generically leads to the emergence of  scale-free production network and distribution of firms' size. The continuous inflow of new firms shifts away the model from a steady state to a ``dynamic equilibrium" in which firms get scaled according to their resistance to competitive forces. Second, the model is able to reproduce a large set of stylized facts of firms' demographics such as the distribution of growth rates or the negative relation between the variance of growth rate and the size of firms.

Disequilibrium and agent-based models are often criticized for their lack of clarity. We arguably overcome this issue here by providing both numerical simulations and analytical results (in an approximation of the model). More broadly, the paper shows that disequilibrium models can provide very clearcut economic intuitions: the parsimony of our model allows to put forward very clearly the relationship between the emergence of fat tails and the asymmetric effects of competition on firms of different size. In contrast, general equilibrium models of industrial dynamics usually have a fairly complex architecture that does not lend itself to a straightforward identification of causal relationships. A major conceptual difference between both approaches is that in general equilibrium competition is a driving force towards equilibrium while, in our framework, competition affects the structure of the economy and disrupts equilibrium.  

The introduction of disequilibrium also sheds new light on the issue of network formation.  Our disequilibrium approach provides simple micro-foundations for the emergence of scale-free networks, while game-theoretic models give raise to much more regular networks than those observed empirically. As a matter of fact, in our setting, if the model were to reach a micro steady-state, scale-free networks would not materialize.    

As a whole, these results illustrate the potential of out-of-equilibrium dynamics to model the formation of production networks while accounting for empirical regularities. A natural extension of this work is to consider the possibility of endogenous technological change and its impact on the structure of the network. Indeed, an originality of our approach is to consider that technology is embedded within the production network. This suggests to consider increasing connectivity in the network as a natural avenue to represent growth through increasing variety. Relatedly, the model could be extended to analyze the impact of policy or environmental shocks on industrial organization. From a more theoretical perspective, our approach could be extended to investigate the formation of other types of socio-economic networks using context-specific models and out-of-equilibrium dynamics in complement to equilibrium game-theoretic approaches that provide very clear intuitions about the determinants of network formation, but are unable to reproduce the complexity of actual networks.

\newpage
   
\bibliographystyle{plainnat}
\bibliography{AgentBased,ABMoney2,ABMoney}

\begin{thebibliography}{43}
\providecommand{\natexlab}[1]{#1}
\providecommand{\url}[1]{\texttt{#1}}
\expandafter\ifx\csname urlstyle\endcsname\relax
  \providecommand{\doi}[1]{doi: #1}\else
  \providecommand{\doi}{doi: \begingroup \urlstyle{rm}\Url}\fi

\bibitem[Acemoglu et~al.(2012)Acemoglu, Carvalho, Ozdaglar, and
  Tahbaz-Salehi]{Acemoglu}
Daron Acemoglu, Vasco~M. Carvalho, Asuman Ozdaglar, and Alireza Tahbaz-Salehi.
\newblock The network origins of aggregate fluctuations.
\newblock \emph{Econometrica}, 80\penalty0 (5):\penalty0 1977--2016, 09 2012.
\newblock URL
  \url{http://ideas.repec.org/a/ecm/emetrp/v80y2012i5p1977-2016.html}.

\bibitem[Amaral et~al.(1997)Amaral, Buldyrev, Havlin, Maass, Salinger, Stanley,
  and Stanley]{amaral}
Luis A~Nunes Amaral, Sergey~V Buldyrev, Shlomo Havlin, Philipp Maass, Michael~A
  Salinger, H~Eugene Stanley, and Michael~HR Stanley.
\newblock Scaling behavior in economics: the problem of quantifying company
  growth.
\newblock \emph{Physica A: Statistical Mechanics and its Applications},
  244\penalty0 (1):\penalty0 1--24, 1997.

\bibitem[Aoki and Yoshikawa(2011)]{aoki2011}
Masanao Aoki and Hiroshi Yoshikawa.
\newblock \emph{Reconstructing macroeconomics: a perspective from statistical
  physics and combinatorial stochastic processes}.
\newblock Cambridge University Press, 2011.

\bibitem[Arrow et~al.(1959)Arrow, Block, and Hurwicz]{arrow1959stability}
Kenneth~J Arrow, Henry~D Block, and Leonid Hurwicz.
\newblock On the stability of the competitive equilibrium, ii.
\newblock \emph{Econometrica: Journal of the Econometric Society}, pages
  82--109, 1959.

\bibitem[Arthur(1994)]{arthur1994}
W~Brian Arthur.
\newblock \emph{Increasing returns and path dependence in the economy}.
\newblock University of Michigan Press, 1994.

\bibitem[Arthur(2006)]{arthur2006out}
W~Brian Arthur.
\newblock Out-of-equilibrium economics and agent-based modeling.
\newblock \emph{Handbook of computational economics}, 2:\penalty0 1551--1564,
  2006.

\bibitem[Axtell(2001)]{axtell2001zipf}
Robert~L Axtell.
\newblock Zipf distribution of us firm sizes.
\newblock \emph{Science}, 293\penalty0 (5536):\penalty0 1818--1820, 2001.

\bibitem[Bak et~al.(1993)Bak, Chen, Scheinkman, and Woodford]{bak}
Per Bak, Kan Chen, Jos{\'e} Scheinkman, and Michael Woodford.
\newblock Aggregate fluctuations from independent sectoral shocks:
  self-organized criticality in a model of production and inventory dynamics.
\newblock \emph{Ricerche Economiche}, 47\penalty0 (1):\penalty0 3--30, 1993.

\bibitem[Barab{\'a}si and Albert(1999)]{barabasi1999}
Albert-L{\'a}szl{\'o} Barab{\'a}si and R{\'e}ka Albert.
\newblock Emergence of scaling in random networks.
\newblock \emph{science}, 286\penalty0 (5439):\penalty0 509--512, 1999.

\bibitem[Barab{\'a}si et~al.(2009)]{barabasi2009}
Albert-L{\'a}szl{\'o} Barab{\'a}si et~al.
\newblock Scale-free networks: a decade and beyond.
\newblock \emph{science}, 325\penalty0 (5939):\penalty0 412, 2009.

\bibitem[Battiston et~al.(2007)Battiston, Delli~Gatti, Gallegati, Greenwald,
  and Stiglitz]{BattistonJEDC}
Stefano Battiston, Domenico Delli~Gatti, Mauro Gallegati, Bruce Greenwald, and
  Joseph~E. Stiglitz.
\newblock Credit chains and bankruptcy propagation in production networks.
\newblock \emph{Journal of Economic Dynamics and Control}, 31\penalty0
  (6):\penalty0 2061--2084, June 2007.
\newblock URL
  \url{http://ideas.repec.org/a/eee/dyncon/v31y2007i6p2061-2084.html}.

\bibitem[Bonart et~al.(2014)Bonart, Bouchaud, Landier, and Thesmar]{bonart2014}
Julius Bonart, Jean-Philippe Bouchaud, Augustin Landier, and David Thesmar.
\newblock Instabilities in large economies: aggregate volatility without
  idiosyncratic shocks.
\newblock \emph{Journal of Statistical Mechanics: Theory and Experiment},
  2014\penalty0 (10):\penalty0 P10040, 2014.

\bibitem[Bottazzi and Secchi(2006)]{bottazzi2006}
Giulio Bottazzi and Angelo Secchi.
\newblock Explaining the distribution of firm growth rates.
\newblock \emph{The RAND Journal of Economics}, 37\penalty0 (2):\penalty0
  235--256, 2006.

\bibitem[Bottazzi et~al.(2001)Bottazzi, Dosi, Lippi, Pammolli, and
  Riccaboni]{bottazzi2001}
Giulio Bottazzi, Giovanni Dosi, Marco Lippi, Fabio Pammolli, and Massimo
  Riccaboni.
\newblock Innovation and corporate growth in the evolution of the drug
  industry.
\newblock \emph{International Journal of Industrial Organization}, 19\penalty0
  (7):\penalty0 1161--1187, 2001.

\bibitem[Carvalho and Voigtl{\"a}nder(2014)]{carvalho2014input}
Vasco~M Carvalho and Nico Voigtl{\"a}nder.
\newblock Input diffusion and the evolution of production networks.
\newblock Technical report, National Bureau of Economic Research, 2014.

\bibitem[Clauset et~al.(2009)Clauset, Shalizi, and Newman]{clauset2009}
Aaron Clauset, Cosma~Rohilla Shalizi, and Mark~EJ Newman.
\newblock Power-law distributions in empirical data.
\newblock \emph{SIAM review}, 51\penalty0 (4):\penalty0 661--703, 2009.

\bibitem[Coad(2009)]{coad2009growth}
Alex Coad.
\newblock \emph{The growth of firms: A survey of theories and empirical
  evidence}.
\newblock Edward Elgar Publishing, 2009.

\bibitem[Dixit and Stiglitz(1977)]{dixit1977}
Avinash~K Dixit and Joseph~E Stiglitz.
\newblock Monopolistic competition and optimum product diversity.
\newblock \emph{The American Economic Review}, pages 297--308, 1977.

\bibitem[Ericson and Pakes(1995)]{ericson1995markov}
Richard Ericson and Ariel Pakes.
\newblock Markov-perfect industry dynamics: A framework for empirical work.
\newblock \emph{The Review of Economic Studies}, 62\penalty0 (1):\penalty0
  53--82, 1995.

\bibitem[Ethier(1982)]{ethier1982}
Wilfred~J Ethier.
\newblock National and international returns to scale in the modern theory of
  international trade.
\newblock \emph{The American Economic Review}, pages 389--405, 1982.

\bibitem[Fisher(2011)]{fisher2011stability}
F~Fisher.
\newblock The stability of general equilibrium: What do we know and why is it
  important.
\newblock \emph{General equilibrium analysis: A century after Walras}, pages
  34--45, 2011.

\bibitem[Fisher(1989)]{fisher1989disequilibrium}
Franklin~M Fisher.
\newblock \emph{Disequilibrium foundations of equilibrium economics}.
\newblock Number~6. Cambridge University Press, 1989.

\bibitem[Gabaix(1999)]{gabaixzipf}
Xavier Gabaix.
\newblock Zipf's law for cities: an explanation.
\newblock \emph{Quarterly journal of Economics}, pages 739--767, 1999.

\bibitem[Gabaix(2009)]{gabaixpower}
Xavier Gabaix.
\newblock Power laws in economics and finance.
\newblock \emph{Annual Review of Economics}, 1\penalty0 (1):\penalty0 255--294,
  2009.

\bibitem[Gualdi et~al.(2015{\natexlab{a}})Gualdi, Bouchaud, Cencetti, Tarzia,
  and Zamponi]{GualBouPRL}
Stanislao Gualdi, Jean-Philippe Bouchaud, Giulia Cencetti, Marco Tarzia, and
  Francesco Zamponi.
\newblock Endogenous crisis waves: Stochastic model with synchronized
  collective behavior.
\newblock \emph{Phys. Rev. Lett.}, 114:\penalty0 088701, Feb
  2015{\natexlab{a}}.
\newblock \doi{10.1103/PhysRevLett.114.088701}.
\newblock URL \url{http://link.aps.org/doi/10.1103/PhysRevLett.114.088701}.

\bibitem[Gualdi et~al.(2015{\natexlab{b}})Gualdi, Tarzia, Zamponi, and
  Bouchaud]{gualdi2015tipping}
Stanislao Gualdi, Marco Tarzia, Francesco Zamponi, and Jean-Philippe Bouchaud.
\newblock Tipping points in macroeconomic agent-based models.
\newblock \emph{Journal of Economic Dynamics and Control}, 50:\penalty0 29--61,
  2015{\natexlab{b}}.

\bibitem[Hopenhayn(1992)]{hopenhayn1992entry}
Hugo~A Hopenhayn.
\newblock Entry, exit, and firm dynamics in long run equilibrium.
\newblock \emph{Econometrica: Journal of the Econometric Society}, pages
  1127--1150, 1992.

\bibitem[Jackson and Rogers(2007)]{jackson2007meeting}
Matthew~O Jackson and Brian~W Rogers.
\newblock Meeting strangers and friends of friends: How random are social
  networks?
\newblock \emph{The American economic review}, pages 890--915, 2007.

\bibitem[Jackson et~al.(2008)]{jackson2008}
Matthew~O Jackson et~al.
\newblock \emph{Social and economic networks}, volume~3.
\newblock Princeton University Press Princeton, 2008.

\bibitem[Kalecki(1945)]{kalecki1945}
Michael Kalecki.
\newblock On the gibrat distribution.
\newblock \emph{Econometrica: Journal of the Econometric Society}, pages
  161--170, 1945.

\bibitem[Klepper and Thompson(2006)]{klepper2006submarkets}
Steven Klepper and Peter Thompson.
\newblock Submarkets and the evolution of market structure.
\newblock \emph{The RAND Journal of Economics}, 37\penalty0 (4):\penalty0
  861--886, 2006.

\bibitem[Klette and Kortum(2004)]{klette2004innovating}
Tor~Jakob Klette and Samuel Kortum.
\newblock Innovating firms and aggregate innovation.
\newblock \emph{Journal of Political Economy}, 112\penalty0 (5):\penalty0
  986--1018, 2004.

\bibitem[K{\"o}nig et~al.(2012)K{\"o}nig, Battiston, Napoletano, and
  Schweitzer]{konig2012}
Michael~D K{\"o}nig, Stefano Battiston, Mauro Napoletano, and Frank Schweitzer.
\newblock The efficiency and stability of r\&d networks.
\newblock \emph{Games and Economic Behavior}, 75\penalty0 (2):\penalty0
  694--713, 2012.

\bibitem[K{\"o}nig et~al.(2014)K{\"o}nig, Tessone, and Zenou]{konig2014}
Michael~D K{\"o}nig, Claudio~J Tessone, and Yves Zenou.
\newblock Nestedness in networks: A theoretical model and some applications.
\newblock \emph{Theoretical Economics}, 9\penalty0 (3):\penalty0 695--752,
  2014.

\bibitem[Luttmer(2007)]{luttmer2007selection}
Erzo~GJ Luttmer.
\newblock Selection, growth, and the size distribution of firms.
\newblock \emph{The Quarterly Journal of Economics}, pages 1103--1144, 2007.

\bibitem[Mahadev and Peled(1995)]{mahadev}
Nadimpalli~VR Mahadev and Uri~N Peled.
\newblock \emph{Threshold graphs and related topics}, volume~56.
\newblock Elsevier, 1995.

\bibitem[Romer(1990)]{romer1990}
Paul~M Romer.
\newblock Endogenous technological change.
\newblock \emph{Journal of Political Economy}, 98\penalty0 (5 pt 2), 1990.

\bibitem[Rossi-Hansberg and Wright(2007)]{rossi2007establishment}
Esteban Rossi-Hansberg and Mark~LJ Wright.
\newblock Establishment size dynamics in the aggregate economy.
\newblock \emph{The American Economic Review}, pages 1639--1666, 2007.

\bibitem[Scheinkman and Woodford(1994)]{scheinkman1994}
Jose~A Scheinkman and Michael Woodford.
\newblock Self-organized criticality and economic fluctuations.
\newblock \emph{The American Economic Review}, pages 417--421, 1994.

\bibitem[Schweitzer et~al.(2009)Schweitzer, Fagiolo, Sornette, Vega-Redondo,
  Vespignani, and White]{schweitzer2009}
Frank Schweitzer, Giorgio Fagiolo, Didier Sornette, Fernando Vega-Redondo,
  Alessandro Vespignani, and Douglas~R White.
\newblock Economic networks: The new challenges.
\newblock \emph{Science}, 325\penalty0 (5939):\penalty0 422--425, 2009.

\bibitem[Simon et~al.(1977)Simon, Ijiri, and Simon]{simon1977skew}
Herbert~A Simon, Y~Ijiri, and HA~Simon.
\newblock Skew distributions and the sizes of business firms, 1977.

\bibitem[Weisbuch and Battiston(2007)]{weisbuch2007production}
G{\'e}rard Weisbuch and Stefano Battiston.
\newblock From production networks to geographical economics.
\newblock \emph{Journal of Economic Behavior \& Organization}, 64\penalty0
  (3):\penalty0 448--469, 2007.

\bibitem[Yang and Borland(1991)]{yang1991}
Xiaokai Yang and Jeff Borland.
\newblock A microeconomic mechanism for economic growth.
\newblock \emph{Journal of Political Economy}, pages 460--482, 1991.

\end{thebibliography}

\newpage

\appendix

\section{Asymptotic Analysis}

\subsection{Proof of Proposition $\ref{propns}$}\label{proofns}

\begin{demo}
A steady state consists in vectors of wealths $\tilde{w}$, prices $\tilde{p},$ productions $\tilde{q}$ and in  an adjacency matrix $\tilde{A}$ satisfying the following properties.
\begin{itemize} 
\item First, according to equation $\ref{netchange},$ each firm $i$ 
buys only from the cheapest suppliers (otherwise it would rewire). That is, one has for all $i \in N:$
 \begin{equation}
\max_{\{j \mid \tilde{a}_{i,j}=1\}} \tilde{p}_j \leq \min_{\{k \mid \tilde{a}_{i,k}=0\}}   \tilde{p}_k  \label{pricecomp} 
\end{equation} 
\item Second, firms only differ in terms of their number of suppliers $n_i$ and, at a steady state, the larger the number of suppliers  of a firm, the more productive and cheaper it is. Otherwise, it could adopt the same production technique as any firm with a smaller number of suppliers and improve upon it by diversifying marginally. Therefore, one has for all $i,j \in M:$
 \begin{equation}
n_i > n_j  \Rightarrow \tilde{p}_i < \tilde{p}_j  \label{ncomp} 
\end{equation} 
\item Combining equations $(\ref{pricecomp})$ and $(\ref{ncomp}),$ one gets:
\begin{equation} \min_{\{j \mid \tilde{a}_{i,j}=1\}} n_j \geq \max_{\{k \mid \tilde{a}_{i,k}=0\}}   n_k  \end{equation} 
and therefore
\begin{equation} n_j<n_i \Rightarrow [\forall h \in M  \ \overline{a}_{h,j}=1 \Rightarrow \overline{a}_{h,i}=1].\end{equation}

\end{itemize}
This ends the proof. 

\end{demo}

\subsection{Structure of steady-state networks} \label{stnet}

Let us denote by $M_{\iota}:=\{ f \in M \mid n_f=\iota\}$ the set of firms with exactly $\iota$ suppliers, by $m_{\iota}$ the number of such firms, by $\iota_1 \geq \iota_2 \geq  \cdots \geq \iota_{r}$ the decreasing sequence of $\iota$s for which  $M_{\iota} \not = \emptyset$ and let $\phi_{i}=\sum_{j=1}^{i} m_{\iota_{j}}$ be the total number of firms with more than $\iota_i$ suppliers. Then, at a steady state, each firm in $M_{\iota_i}$ is linked to every firm in the $M_{\iota_k}$s such that $\phi_{\iota_k} \leq \iota_i,$ and to $\iota_i-\phi_{\ell}$ firms in the cluster
$\ell$ such that  $\phi_{\ell} \leq \iota_i- < \phi_{\ell+1}.$ Accordingly, the in-degree distribution of the network is such that all firms in $M_{\iota_i}$ have approximately the same number of incoming links $\sum_{\{k \mid k \geq \iota_i \}}  m_k.$  

\end{document}